\documentclass{aa}  

\usepackage{graphicx}
\usepackage{lscape}
\usepackage{rotating}
\usepackage[labelfont=bf]{caption}
\usepackage{subfig}
\usepackage{amsmath}
\usepackage{natbib}
\usepackage{txfonts}
\usepackage{hyperref}
\hypersetup{colorlinks=true,citecolor=blue,linkcolor=blue}
\newcommand{\change}[1]{#1}
\newcommand{\changetwo}[1]{#1}
\newcommand{\changethree}[1]{#1}
\newcommand{\changeref}[1]{#1}
\pdfoutput=1

\begin{document}

   \title{Upper limits on CH$_3$OH in the HD 163296 protoplanetary disk}

   \subtitle{Evidence for a low gas-phase CH$_3$OH/H$_2$CO ratio}

   \author{M.T. Carney \inst{1}, M.R. Hogerheijde \inst{1}, V.V. Guzm\'{a}n \inst{2,3}, C. Walsh \inst{4}, K.I. \"{O}berg \inst{5}, \\ 
	   E.C. Fayolle \inst{6}, L.I. Cleeves \inst{7}, J.M. Carpenter \inst{2}, C. Qi \inst{5}}

   \institute{Leiden Observatory, Leiden University, PO Box 9513, 2300 RA, The Netherlands. \\
              \email{masoncarney@strw.leidenuniv.nl}
          \and
	     Joint ALMA Observatory (JAO), Alonso de Cordova 3107, Vitacura, Santiago de Chile, Chile
          \and
             Instituto de Astrof\'{i}sica, Ponticia Universidad Cat\'{o}lica de Chile, Av. Vicu\~{n}a Mackenna 4860, 7820436 Macul, Santiago, Chile
          \and   
             School of Physics and Astronomy, University of Leeds, Leeds LS2 9JT, UK
          \and
             Department of Astronomy, Harvard University, Cambridge, MA 02138, USA
          \and
             Jet Propulsion Laboratory, California Institute of Technology, 4800 Oak Grove Drive, Pasadena, CA 91109-8099, USA
          \and
	     University of Virginia, Charlottesville, VA 22904, USA
             }

   \date{Received 01 October 2018; accepted 08 January 2019}

% \abstract{}{}{}{}{} 
% 5 {} token are mandatory
 
  \abstract
   {Methanol (CH$_3$OH) is at the root of organic ice chemistry in protoplanetary disks. Its connection to prebiotic chemistry
   and its role in the chemical environment of the disk midplane makes it an important target for disk chemistry studies.
   \changethree{However, its weak emission has made detections difficult.} 
   To date, gas-phase CH$_3$OH has been detected in \changethree{only one Class II disk, TW Hya.}
%    The CH$_3$OH molecule represents one of the simplest organics in protoplanetary disks. Methanol is therefore of great interest as a tracer 
%    of the early stages leading to prebiotic chemistry and is expected to play a significant role in the chemical environment of the protoplanetary disk midplane.
   }
   {%To date CH$_3$OH has been detected in the protoplanetary disk around TW Hya only. 
   We aim to constrain the methanol content of the HD 163296 protoplanetary disk.
   }
   {
%    We use the Atacama Large Millimeter/submillimeter Array to search for a total of four CH$_3$OH emission lines,
%    two in band 6 and two in band 7, toward the young Herbig Ae star HD 163296. 
%    The lines are analyzed individually, stacked by band, and stacked all together to increase sensitivity.
%    We compare the results of these observations to those from the TW Hya disk, the only current detection of 
%    gas-phase methanol in a protoplanetary disk. The disk-averaged column density of methanol and formaldehyde is estimated 
%    for each disk assuming optically thin emission in local thermodynamic equilibrium, and we compare the methanol-to-formaldehyde 
%    disk-averaged column density ratio between the disks using H$_2$CO observations taken from the literature. 
   \changetwo{We use the Atacama Large Millimeter/submillimeter Array (ALMA) to search for a total of four CH$_3$OH emission lines in bands
   6 and 7 toward the disk around the young Herbig Ae star HD 163296. The disk-averaged column density of methanol and its related 
   species formaldehyde (H$_2$CO) are estimated assuming optically thin emission in local thermodynamic equilibrium. We compare these 
   results to the \changethree{gas-phase column densities of the} TW Hya disk.}
   }
   {No targeted methanol lines were detected with Keplerian masking in the image plane nor with matched filter analysis in the $uv$ plane
   individually \changetwo{or after line stacking. The 3$\sigma$ disk-integrated intensity upper limits}
   are \change{$< 51$ mJy km s$^{-1}$ for the band 6 lines and $< 26$ mJy km s$^{-1}$} for the band 7 lines. The band 7 lines provide the 
   strictest 3$\sigma$ upper limit on disk-averaged column density \changethree{with $N_{\rm avg} < 5.0 \times 10^{11}$ cm$^{-2}$}. 
   The methanol-to-formaldehyde ratio is ${\rm CH_3OH}/{\rm H_2CO}$ $< 0.24$ in the HD 163296 disk compared to a \changethree{ratio of 1.27 in the 
   TW Hya disk.} %\change{and a ratio of $\sim$0.33 in recent T Tauri disk chemical models}.
   }
   {
   The HD 163296 protoplanetary disk is less abundant in methanol with respect to formaldehyde compared to the disk around TW Hya. 
   \changetwo{Differences in the stellar irradiation in this Herbig Ae disk as compared to that 
   of a disk around a T Tauri star likely influence the gaseous methanol and formaldehyde content. Possible reasons for the lower 
   HD 163296 methanol-to-formaldehyde ratio include:} a higher than expected 
   gas-phase formation of H$_2$CO in the \changeref{HD 163296} disk, uncertainties in the grain surface formation efficiency of CH$_3$OH and H$_2$CO, 
   and differences in the disk structure and/or CH$_3$OH and H$_2$CO desorption processes that drive the release of the molecules 
   from ice mantles back into the \changeref{gas phase}. \changethree{These results provide observational evidence that the gas-phase
   chemical complexity found in disks \changeref{may be} strongly influenced by the spectral type of the host star.}
   }

   \authorrunning{Carney, M.~T. et al.}
   \titlerunning{CH$_3$OH upper limits in HD 163296}
   
   \maketitle
%
%________________________________________________________________

\section{Introduction}

Methanol is an \changethree{astrobiologically relevant molecule} because it acts as a precursor
to more complex organic molecules (COMs) that may develop into amino acids
and other building blocks of life \citep{Oberg2009,Herbst2009}. Maintaining an understanding of methanol
chemistry through the numerous stages of star and planet formation is essential to 
make predictions on the molecular complexity available for incorporation into
extrasolar planetary bodies \citep[e.g.,][]{Drozdovskaya2014}. The CH$_3$OH molecule is regularly 
detected in the earlier, embedded stages of star formation both in the solid 
phase through ice absorption \citep{Grim1991,Skinner1992,Dartois1999,Pontoppidan2004,Bottinelli2010,Kristensen2010,Shimonishi2010,Boogert2015} 
and in the \changeref{gas phase} \citep{Friberg1988,vanDishoeck1995,Graninger2016,Lee2017}. 
These observations provide evidence for \changetwo{the presence of CH$_3$OH ices}
in cold \changetwo{molecular clouds and protostellar envelopes}.

To date, methanol has been detected in \changethree{two protoplanetary disks:
the Class II TW Hydrae \citep{Walsh2016b} and the younger Class I V883 Orionis,
an outbursting FU Orionis object \citep{vantHoff2018}.}
% , which is the closest disk 
% to our own solar system at $60.1 \pm 0.1$ pc \citep{Gaia2018}.
\changetwo{There are currently few informative upper limits on gas-phase methanol in disks.}
The reason for the apparent absence of gas-phase methanol in protoplanetary disks is not 
immediately obvious because CH$_3$OH is expected to form via the hydrogenation
of CO ices \citep{Watanabe2003,Cuppen2009} on the surface of dust grains. %deep in the shielded disk midplane. 
\changethree{Also, the colder, outer regions of protoplanetary disks are expected to 
inherit a reservoir of methanol ice formed earlier, during the protostellar or interstellar phase.} 
Methanol is produced by the same grain surface 
formation pathway as formaldehyde, which is readily detected in disks
\citep{Aikawa2003,Oberg2010b,Qi2013,vanderMarel2014,Loomis2015,Oberg2017,Carney2017}.
\changetwo{However, because of \changeref{the much higher methanol binding (desorption) energy
\citep[e.g., $E_{\rm des}$ of $\sim2000$ K for H$_2$CO and $\sim5500$ K for CH$_3$OH
in mixtures of water ice;][]{Collings2004,Garrod2006}, methanol}
is expected to be frozen out over a much larger region of the disk than formaldehyde.}

\changetwo{Variation in} the formaldehyde and methanol content
across protoplanetary disks may point to differences in their formation 
processes. \changetwo{Formaldehyde can be formed in the \changeref{gas phase} and on grain surfaces,
therefore a lower than expected methanol-to-formaldehyde ratio could be due to} a more 
efficient gas-phase pathway to form H$_2$CO \citep{Fockenberg2002,Atkinson2006}, 
\changetwo{less efficient conversion of H$_2$CO into CH$_3$OH on grain surfaces than expected,}
% differences in their respective desorption processes when released from ices back into the gas phase, e.g., 
or lower than expected CH$_3$OH photodesorption rates and/or immediate UV photodissociation of gas-phase CH$_3$OH \citep{Bertin2016,CruzDiaz2016}.
% Further observations of both methanol and formaldehyde \changethree{in protoplanetary disks} will help to 
% constrain our knowledge of the ongoing physics and 
% chemistry of the disk midplane and \changethree{enable the quantification of the relative importance}
% of grain surface reactions and photodesorption.

% I think this could be said more clearly. I.e. we know that H2CO forms both in the gas-phase and on 
% the grain surface. A higher than expected H2CO/CH3Oh ratio could be due to 1. A more efficient 
% H2CO gas-phase formation than expected. 2. A lower than expected conversion of H2CO to CH3OH on 
% grains because of e.g. H starvation, 3. a lower than expected CH3OH desorption.

\begin{table*}[!t]
 \caption{HD 163296 observational parameters}
 \centering
 \label{tab:obs_par}
 \begin{tabular}{lcc}
%  \hline \hline
%  \multicolumn{3}{c}{Project 2016.1.00884.S} \\
 \hline \hline
 \multicolumn{3}{c}{\vspace{0.2cm} \textbf{Band 6}} \\
 Dates Observed & \multicolumn{2}{c}{2016 November 11, December 01; 2017 March 15 } \\
 Baselines & \multicolumn{2}{c}{15 -- 1000 m | 12 -- 776 k${\rm \lambda}$ } \\
 \hline
  & CH$_3$OH 5$_{05}$--4$_{04}$ (E) & CH$_3$OH 5$_{05}$--4$_{04}$ (A) \vspace{0.2cm} \\
 Rest frequency [GHz] & 241.700 & 241.791 \\
 Synthesized beam [FWHM] & $1.46\arcsec \times 1.13\arcsec$ & $1.46\arcsec \times 1.13\arcsec$ \\
 Position angle & $-76.6^\circ$ & $-76.6^\circ$ \\
 Channel width [km s$^{-1}$] & 0.303 & 0.303 \\
 rms noise\tablefootmark{a} [mJy beam$^{-1}$]  & 3.0 & 3.0 \\
 Weighting & natural & natural \\
 \hline
 Continuum frequency [GHz] & \multicolumn{2}{c}{233.0}  \\
 Synthesized beam [FWHM] &  \multicolumn{2}{c}{$0.55\arcsec \times 0.37\arcsec$}  \\
 Position angle &  \multicolumn{2}{c}{$76.8^\circ$} \\
 rms noise [mJy beam$^{-1}$]  & \multicolumn{2}{c}{0.17} \\
 Integrated flux [mJy]  & \multicolumn{2}{c}{754 $\pm$ 75}  \\
 Weighting & \multicolumn{2}{c}{Briggs, robust = 0.5} \\
 \hline
 
 \hline \hline
 \multicolumn{3}{c}{\vspace{0.2cm} \textbf{Band 7}} \\
 Dates Observed & ACA & 2016 October 05, 08, 13, 26 \\
		& \changeref{12-meter} array & 2017 April 13 \\
 Baselines & ACA & 9 -- 49 m | 9 -- 48 k${\rm \lambda}$ \\
	   & \changeref{12-meter} array & 15 -- 460 m | 15 -- 454 k${\rm \lambda}$ \\
 \hline
  & CH$_3$OH 1$_{10}$--1$_{01}$ (A) & CH$_3$OH 2$_{11}$--2$_{02}$ (A) \vspace{0.2cm} \\
 Rest frequency [GHz] & 303.367 & 304.208 \\
 Synthesized beam [FWHM] & $1.37\arcsec \times 1.14\arcsec$ & $1.36\arcsec \times 1.15\arcsec$ \\
 Position angle & 91.1$^\circ$ & 90.6$^\circ$ \\
 Channel width [km s$^{-1}$] & 0.139 & 0.139 \\
 rms noise\tablefootmark{a} [mJy beam$^{-1}$]  & 2.5 & 2.5 \\
 Weighting & natural & natural \\
 \hline
 Continuum frequency [GHz] & \multicolumn{2}{c}{296.0}  \\
 Synthesized beam [FWHM] &  \multicolumn{2}{c}{$0.63\arcsec \times 0.48\arcsec$}  \\
 Position angle &  \multicolumn{2}{c}{87.8$^\circ$} \\
 rms noise [mJy beam$^{-1}$]  & \multicolumn{2}{c}{0.09} \\
 Integrated flux [mJy]  & \multicolumn{2}{c}{1288 $\pm$ 128}  \\
 Weighting & \multicolumn{2}{c}{Briggs, robust = 0.5} \\

 \hline

 \end{tabular}
 \tablefoot{Flux calibration accuracy is taken to be 10\%. For specifics on the line transition data, see Table~\ref{tab:ch3oh_col_dens_abun}.
 \tablefoottext{a}{Noise levels are per image channel.}
 }
\end{table*}

The HD 163296 (MWC 275) system is an ideal testbed for exploring chemical processing in
protoplanetary disks, in particular for organics. It is
an isolated Herbig Ae pre-main sequence (PMS) star with spectral type A2Ve at an age of
$\sim$5 Myr \citep{Alecian2013}. The star is surrounded by a large, bright protoplanetary 
disk \changeref{containing a significant reservoir of gas} that extends out to
$\sim$550 AU in the gas based on CO measurements \citep{deGregorioMonsalvo2013}. The disk
has an inclination of 44$^{\circ}$, a position angle of 133$^\circ$,
and a total mass of $M_{\rm disk} \approx 0.09 M_{\odot}$
based on physical models \citep{Qi2011,Rosenfeld2013}. 
At such an inclination, the vertical structure as well as the radial structure
can be inferred directly from the molecular line emission maps \citep{Rosenfeld2013,Flaherty2015}. 

\change{Recent measurements of the stellar parallax by Gaia put the HD 163296 system %$9.85 \pm 0.11$ mas, 
at a distance of $d = 101 \pm 1$ pc \citep{Gaia2018}, 
significantly closer than previous distance estimates of 122 pc \citep{vandenAncker1998}.\footnote{The 
updated distance $d = 101 \pm 1$ pc results in a stellar luminosity 
% decrease of $\sim$0.69 compared to previous estimates of 33 L$_{\odot}$ \citep{Alecian2013}.
\changetwo{of $\sim$23 L$_{\odot}$, which is 30\% lower than the previous estimate \citep{Alecian2013}.}
Applying the adjusted luminosity value to the H-R diagram used by 
\citet{Alecian2013} to determine the age of the system and stellar mass results in
an updated age of $\sim$9 Myr and an adjusted stellar mass closer to 2.1 M$_{\odot}$.}
While the new distance will
affect the stellar parameters, this work adopts the previously reported values for 
stellar mass \citep[2.3 M$_{\odot}$;][]{Qi2011} and distance ($d = 122$ pc).
\changethree{The analysis presented here focuses on the disk-averaged molecular column density ratios
of methanol and formaldehyde within the same disk.
The column density is derived from 
the disk-integrated line flux, therefore the updated Gaia distance measurements will affect the line flux similarly 
for molecular species within the same disk, and the effect of the new distance is canceled out.}}

The proximity and size of the disk combined with \changeref{the high total luminosity of the}
Herbig Ae PMS star provides a unique opportunity to fully resolve the
location of the CO snow line, i.e., the midplane radius beyond which gas-phase CO will 
freeze out into ice \citep{Qi2011,Mathews2013,Qi2015}.
Current estimates by \citet{Qi2015} place the CO snow line at a midplane
radius of 90 AU, corresponding to a gas \changethree{and dust} temperature of $\sim$24 K in this disk.
\change{Recent work has revealed that the disk consists of several 
rings and gaps in the millimeter dust and in the gas \citep{Isella2016},
while the CO gas shows asymmetries at specific
velocities \citep{Pinte2018,Teague2018}, both of which may be indicative 
of planet-disk interaction from embedded forming planets.}
Given its large radial extent of $\sim$550 AU and resolved, relatively close-in 
CO snow line position, HD 163296 is one of the best candidates to probe the formation of
organics that require the freeze-out of abundant volatiles such as CO. 

This paper presents observations from the Atacama Large Millimeter/submillimeter Array
(ALMA) of the CH$_3$OH molecule toward HD 163296. Section~\ref{sec:obs} describes the observations
and data reduction. Results including the 
upper limits on the methanol content of the HD 163296 disk and a comparison 
to the TW Hya disk are described in Section~\ref{sec:res}. 
In Section~\ref{sec:disc} we discuss the implications of the
upper limits on the detectability of methanol in disks similar to HD 163296.
Section~\ref{sec:concl} presents the conclusions of this work.

%__________________________________________________________________

\section{Observations and reduction}
\label{sec:obs}

HD 163296 (J2000: R.A. = 17$^{\rm{h}}$56$^{\rm{m}}$21.280$^{\rm{s}}$, 
DEC = --21$^\circ$57$\arcmin$22.441$\arcsec$) was observed
with ALMA in band 6 and band 7 during Cycle 4 under project 2016.1.00884.S. 
Band 6 and band 7 are
receivers operating in the 211--275 GHz and 275--373 GHz range, respectively.
Band 6 observations were done with the ALMA 12-meter array on 
2016 November 11, 2016 December 01, and 2017 March 15 with 42 antennas.
Band 7 observations were carried out with the Atacama Compact Array (ACA) 
on 2016 October 05, 08, 13, 26 using 10 of the \changeref{7-meter} ACA antennas,
and with the ALMA 12-meter array on 2017 April 13 using 45 antennas. 
% Baselines ranged from 15 -- 1445 m in band 6 and from 9 -- 460 m in 
% band 7 to achieve a spatial resolution of $\sim$0.5$\arcsec$. 
In total, four transitions of CH$_3$OH were \changeref{targeted} across the two bands
with the frequency domain mode (FDM) correlator setting:
two CH$_3$OH 5$_{05}$--4$_{04}$ (A/E) lines\footnote{\changeref{As a methyl group molecule, methanol exists in three forms with different 
hydrogen spin symmetry properties. The A-type form has a total spin $3/2$, while the E-type form is 
degenerate having E$^{\rm a}$ and E$^{\rm b}$ varieties with total spin $1/2$.
The ratio of A-type to E-type forms of methanol is one.}} in band 6 at 241.791 GHz and 241.700 GHz
with a frequency (velocity) resolution of 244 kHz (0.303 km s$^{-1}$); 
and in band 7, CH$_3$OH 2$_{11}$--2$_{02}$ (A) at 304.208 GHz and CH$_3$OH 1$_{10}$--1$_{01}$ (A)
at \changeref{303.367 GHz} with a frequency (velocity) resolution of 141 kHz (0.139 km s$^{-1}$).
All CH$_3$OH lines were in the upper side band (USB) of their 
execution blocks. The lower side band (LSB) contained observations of the continuum,
C$^{17}$O $J=2-1$, CN $J=2-1$, and CH$_3$CN $J=13-12$ in band 6, and the continuum, DCN $J=4-3$, and four H$_2$CO lines in band 7
\changeref{which will be presented in Guzm\'{a}n, et al. (in prep)}.
Table~\ref{tab:obs_par} summarizes the observational parameters for each CH$_3$OH line and the continuum.

\changethree{Band 6 observations} were obtained over three execution blocks
% of 51 min ($\times$2) and 38 min ($\times$1) 
% at 6.05 seconds per integration step for a total time on-source of 68 minutes. 
with 6.05 sec integration steps and 68 minutes \changethree{total time} on-source.
System temperatures varied from 60--140 K and the average precipitable water vapor
varied from 1.5--2.3 mm.
% The average precipitable water vapor was 1.5 mm on 2016 November 11,
% 2.3 mm on 2016 December 1, and 2.1 mm on 2017 March 15.
J1924-2914 was the bandpass calibrator and Titan was the flux 
calibrator for all execution blocks. \changeref{The average flux values for Titan were:
1.15 Jy in the USB and 1.01 Jy in the LSB for 2016 November 11 and December 01;
0.963 Jy in the USB and 0.846 Jy in the LSB for 2017 March 15.}
The gain calibrator was different
for each execution block: J1745-2900 on 2016 November 11, J1742-1517 on 
2016 December 01, and J1733-1304 on 2017 March 15. The \changeref{derived} flux values 
for J1745-2900, J1742-1517, and J1733-1304 were 3.29 Jy, \changeref{0.212 Jy}, and 
1.47 Jy, respectively. All measurement sets were subsequently 
concatenated and time binned to 30s integration time per visibility 
for imaging and analysis.

\changethree{Band 7 observations} were obtained with the 12-meter array
over three execution blocks %of 78 min ($\times$3) 
% at 6.05 seconds per integration step for a total time on source of 105 minutes and
with 6.05 sec integration steps and 105 minutes \changethree{total time} on-source.
Data was also obtained with the ACA over four execution blocks %of 119 min ($\times$4) 
% at 10.1 seconds per integration step for a total time on-source of 184 minutes.
with 10.1 sec integration steps and 184 minutes \changethree{total time} on-source.
System temperatures varied from 80--150 K and the average precipitable water vapor
varied from 0.5--1.1 mm.
% The average precipitable water vapor was 0.5 mm on 2016 October 26,
% 1.1 mm on 2016 October 08, and 0.8 mm on 2016 October 05, 13 and 2017 April 13.
J1924-2914 was the bandpass calibrator for all execution blocks. Titan, Neptune,
J1733-1304, and J1751+0939 were used as flux calibrators. 
\changeref{The average flux values were: Titan -- 1.96 Jy in the USB and 1.82 Jy in the LSB for 2017 April 13;
Neptune -- 22.5 Jy in the USB and 21.2 Jy in the LSB for 2016 October 08, 26;
J1733-1304 -- 1.32 Jy (2017 April 13), 1.14 Jy (2016 October 13) in the USB and 
1.36 Jy (2017 April 13), 1.18 Jy (2016 October 13) in the LSB; 
J1751${\rm +}$0939 -- 1.58 Jy in the USB and 1.60 Jy in the LSB for 2016 October 05.} The gain calibrators 
were J1733-1304 for the \changeref{12-meter} array data and J1745-2900 for the ACA data. 
The \changeref{derived} flux value for J1733-1304 was 1.36 Jy (2017 April 13) and the values for 
\changeref{J1745-2900} were 3.2 Jy (2016 October 05, 08, 13), and 4.6 Jy (2016 October 26). 
All measurement sets were subsequently concatenated and time binned to 
30 sec integration time per visibility for imaging and analysis.

Self-calibration for HD 163296 in band 6 was done with five spectral
windows dedicated to continuum observations: two in the LSB
at 223.5 GHz and 224 GHz and three in the USB at 234 GHz, 
241 GHz, and 242 GHz with a total combined bandwidth of 469 MHz.
The band 6 reference antenna was DA41.
Band 7 self-calibration was done with three spectral windows
dedicated to continuum observations: one in the LSB
at 289 GHz and two in the USB at 302 GHz and 303.5 GHz 
with a total combined bandwidth of 469 MHz.
The band 7 reference antenna was DA59 for the \changeref{12-meter} array and 
CM03 for the ACA.
A minimum of four baselines per antenna and a minimum
signal-to-noise ratio (SNR) of two were required for self-calibration. 
Calibration solutions were calculated twice for phase and once for amplitude.
The first phase solution interval (solint) was 200 sec, the second
phase and amplitude solutions had solint equal to the binned
integration time (30 sec). Self-calibration solutions for the continuum
spectral windows were mapped to the \changeref{line spectral windows nearest}
in frequency. Continuum subtraction for the line data was done 
in the $uv$ plane using a single-order polynomial fit to the line-free channels.
CLEAN imaging was performed with natural weighting for each continuum-subtracted
CH$_3$OH line with a $uv$ taper 
to achieve a 1$\arcsec$ beam in order to increase the sensitivity.

This paper also makes use of \changethree{Submillimeter Array (SMA) H$_2$CO data for the HD 163296 disk
\citep{Qi2013},} ALMA H$_2$CO data for the HD 163296 disk \citep{Carney2017}, 
ALMA H$_2$CO data for the TW Hya disk \citep{Oberg2017},
and ALMA CH$_3$OH data for the TW Hya disk \citep{Walsh2016b}. 
The following software and coding languages are used for 
data analysis: the \textsc{casa} package \changetwo{version 4.7.2} \citep{McMullin2007}
% the \textsc{miriad} package \citep{Sault1995}, 
and \textsc{python}.

%______________________________________________________________

\section{Results}
\label{sec:res}

\begin{figure}[!t]
 \centering
 \includegraphics[width=0.45\textwidth]{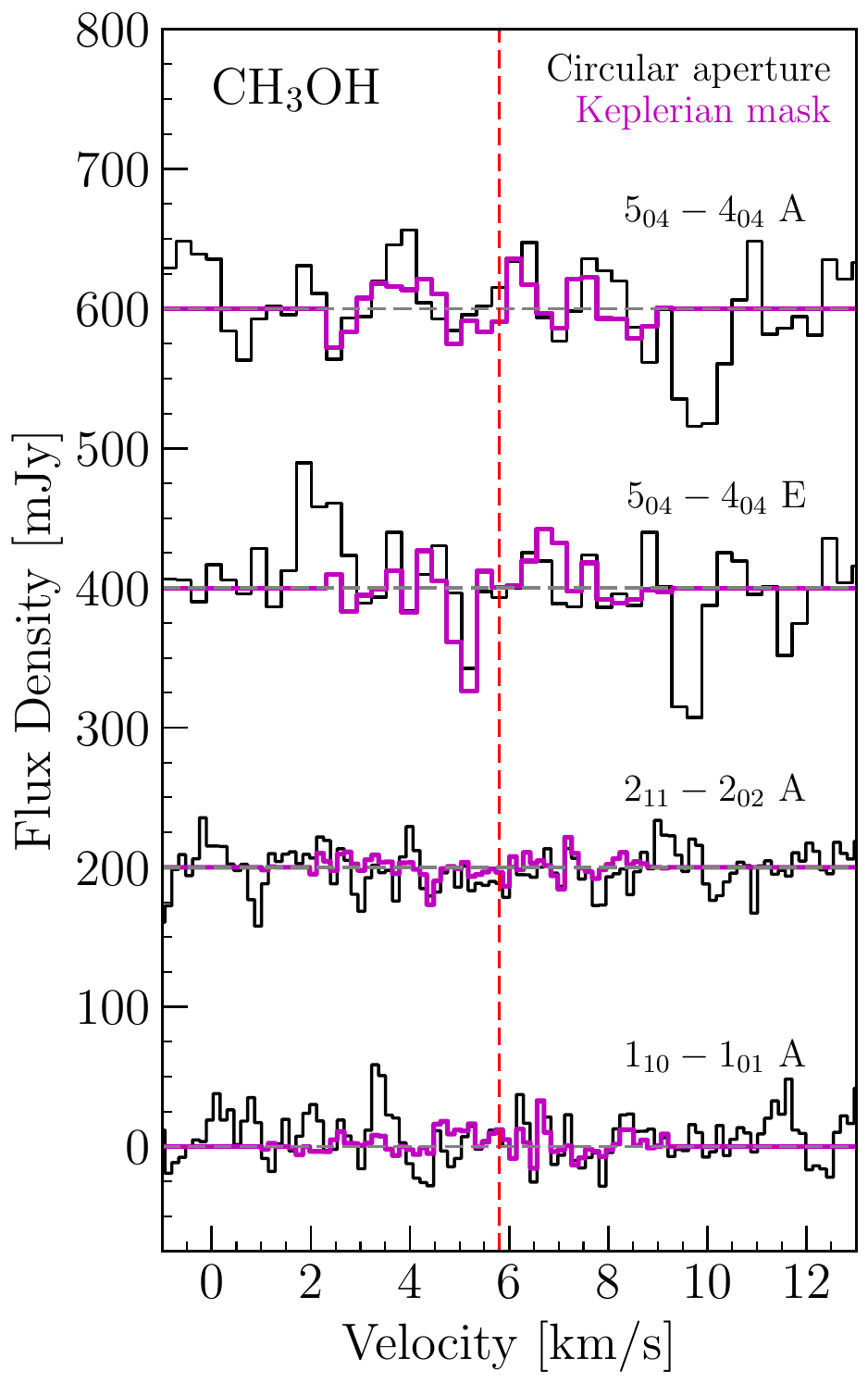}
 \caption{
 Spectra \changeref{at the expected velocity} of CH$_3$OH \changeref{line emission} in the HD 163296 
 disk showing non-detections from 
 aperture-masked image cubes using an 8$\arcsec$ diameter
 circular aperture (black) and Keplerian-masked image cubes (magenta). 
 The two bottom spectra are observed in band 7 in 0.139 km s$^{-1}$ channels while the 
 two top spectra are in band 6 in 0.303 km s$^{-1}$ channels. The horizontal
 gray dashed line represents the spectrum baseline, which is offset
 by 200 mJy for each line. The vertical red dashed line shows
 the systemic velocity at 5.8 km s$^{-1}$ \citep{Qi2011}.
 }
 \label{fig:spectra}
\end{figure}

No methanol lines listed in Table~\ref{tab:obs_par} are detected
in the disk around HD 163296 neither individually nor after line stacking. 
\changetwo{In this section, we first describe the stacking and masking methods used
to maximize the SNR to attempt to extract the 
disk-integrated intensity of the CH$_3$OH lines. The method used to estimate 
the column density and abundance of methanol in the HD 163296 disk
is then described. A comparison is presented between the CH$_3$OH 
and H$_2$CO content in the disks around HD 163296 and TW Hya based 
on data taken from the literature. Finally, model spectra of the 
band 7 CH$_3$OH lines are created for HD 163296 and compared to 
the sensitivity of the observations.}

\subsection{Line extraction}
\label{sec:res_lineex}

\changetwo{We attempt to extract the targeted CH$_3$OH lines from the CLEANed image 
cubes using a circular aperture with an 8$\arcsec$ \changeref{diameter} centered on the source,
which yields no detections (see Figure~\ref{fig:spectra}).
To increase the SNR we repeat this analysis after stacking the CH$_3$OH lines using different 
line stacking schemes.}
We further attempt to increase the SNR
of the CH$_3$OH data by applying masking techniques: 
Keplerian masking in the image plane, and matched filter analysis 
in the $uv$ plane to search for any signal in the raw visibilities.

\subsubsection{Line stacking}
\label{sec:res_linestack}

Stacking is done for band 6 and band 7 lines separately, and then again 
for \changethree{both bands} together. The band 7 lines are more easily excited
due to their lower upper energy ($E_{\rm u} < 22$ K) values compared 
to the band 6 lines ($E_{\rm u} > 34$ K; see Table~\ref{tab:ch3oh_col_dens_abun}),
\changetwo{thus band 7 observations should be sensitive to lower
CH$_3$OH column densities and should be easier to detect.}
\changeref{Note that the level populations are likely to be in LTE for 
the expected methanol emitting region where gas densities in the disk are high ($\gtrsim10^7$ cm$^{-3}$), 
thus we do not expect the critical density of the lines to influence the 
amount of line emission (see Table~\ref{tab:ch3oh_col_dens_abun}).}

\changetwo{First, we stack \changethree{the lines} in the image plane by adding together the integrated intensity 
maps ($v = 2.4 - 9.2$ km s$^{-1}$) created from the CH$_3$OH continuum-subtracted 
and $uv$-tapered CLEANed image cubes. 
Second, we stack in the $uv$ plane by concatenating ALMA measurement sets
prior to imaging. Stacking in the $uv$ plane is done using the
\textsc{casa} \texttt{cvel} function, which is used to regrid the velocity axis of 
line data and has the option to combine visibility data for multiple lines.
For $uv$ stacking across all bands, the band 7 lines are regridded 
to 0.303 km s$^{-1}$ channels to match the channel width of the band 6 lines.}
\changetwo{Methanol remains undetected after implementing the stacking 
methods described above. }

\subsubsection{Keplerian masking in the image plane}
\label{sec:res_immask}

\begin{figure*}[!t]
 \centering
 \includegraphics[width=0.33\textwidth]{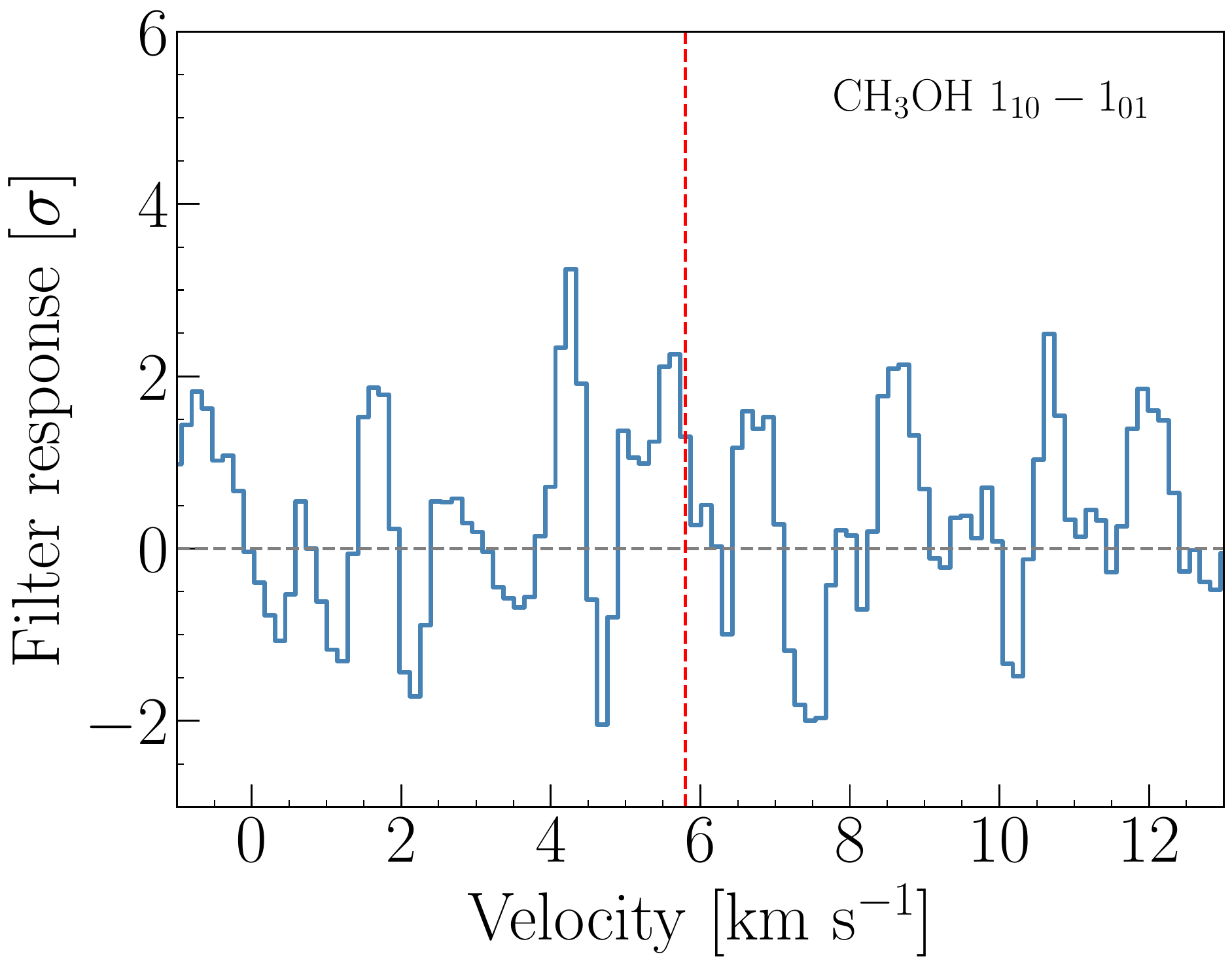} 
 \includegraphics[width=0.33\textwidth]{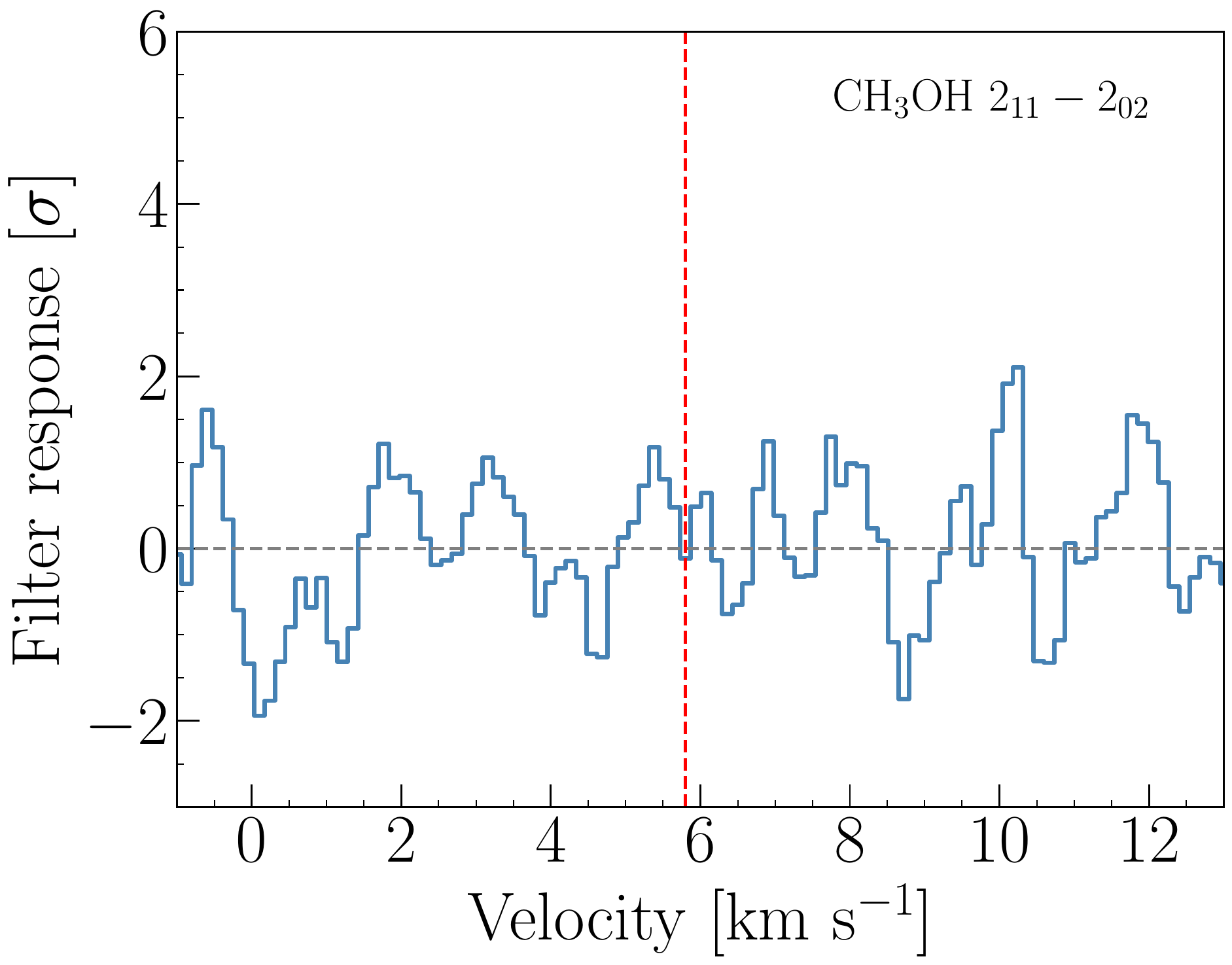} 
 \includegraphics[width=0.33\textwidth]{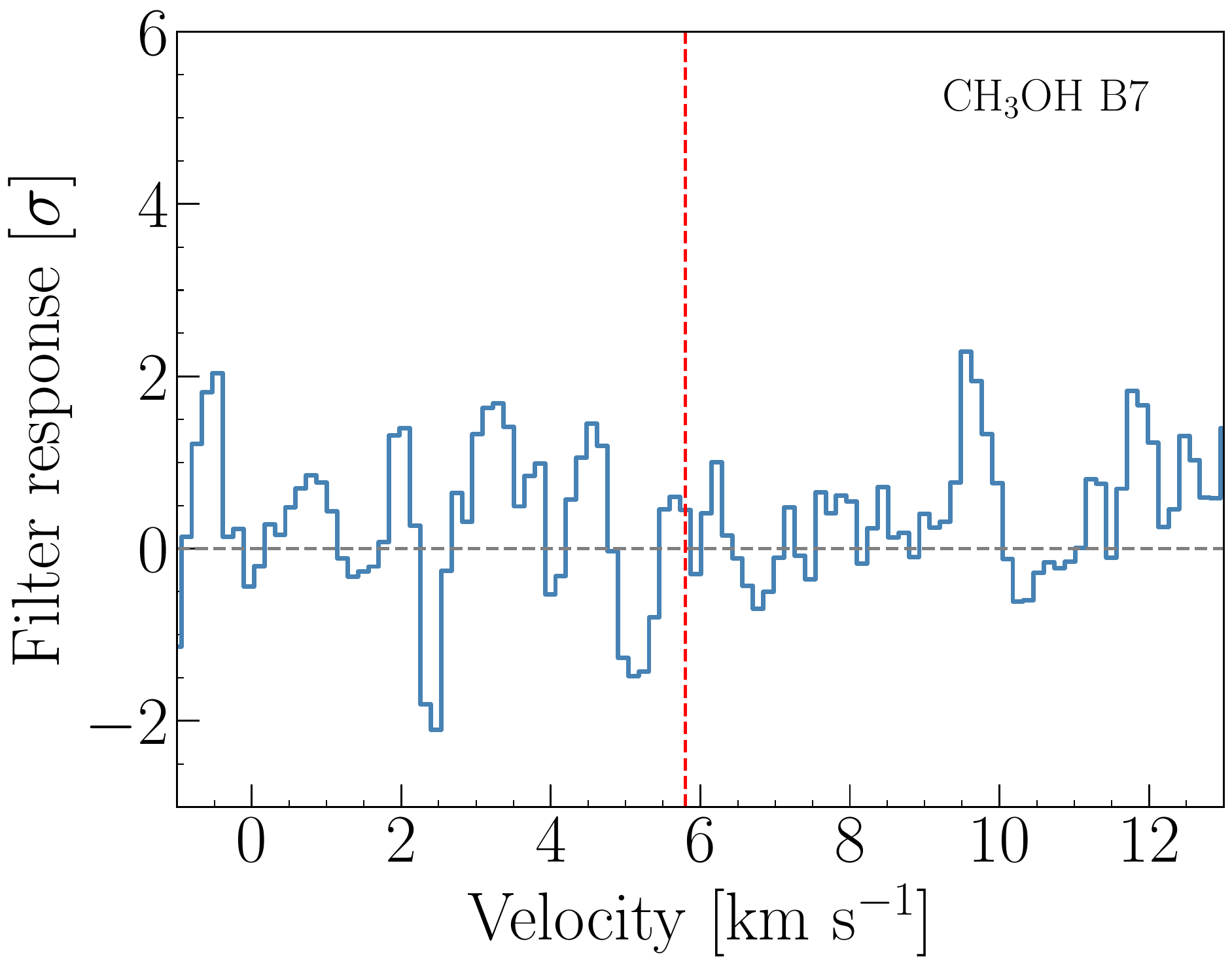}
 \caption{Matched filter results for the band 7 CH$_3$OH lines in the HD 163296 disk using the H$_2$CO emission as a template.
 A peak $>3\sigma$ at the source velocity ($v=5.8$ km s$^{-1}$; red dashed line) would signify a positive detection of methanol.
 The band 7 lines should be the strongest in our sample, but there is no 
 evidence of CH$_3$OH in the matched filter for the band 7 individual lines nor 
 the band 7 stacked lines.
 (Left) CH$_3$OH 1$_{10}$--1$_{01}$ (A) line. (Middle) CH$_3$OH 2$_{11}$--2$_{02}$ (A) line. 
 (Right) Stacked band 7 CH$_3$OH lines. 
 }
 \label{fig:matchedfilter}
\end{figure*}

For maximum SNR in the image plane, we apply a Keplerian mask to the CLEANed image
cube for each CH$_3$OH line \citep{Carney2017,Salinas2017} to exclude 
noisy pixels that are not associated with the \changethree{emission expected from a disk in Keplerian rotation}. The mask is based on the 
velocity profile of a rotating disk, which is assumed to be Keplerian around a central stellar 
mass of $M = 2.3 \ M_{\odot}$ \citep{Alecian2013}. \change{A subset of pixels ($x,y,v$) are identified
in the CH$_3$OH image cubes where the Doppler-shifted line velocity \changetwo{projected along the line of sight}
matches the pixel Keplerian velocity ($x,y,v_{\rm K}$) \changetwo{projected along the line of sight.} 
Pixels with velocities that do not match the Keplerian rotational profile criteria are masked.} 
\change{Integrated intensity maps and disk-integrated spectra are again created from 
the Keplerian-masked cubes} of the CH$_3$OH lines individually and after line stacking; 
\changetwo{however, in all cases, CH$_3$OH remains undetected.}
Figure~\ref{fig:spectra}
shows the aperture-masked spectra and the Keplerian-masked spectra of the four
methanol lines \changeref{targeted} in HD 163296.

Upper limits on the integrated intensity for each CH$_3$OH line are calculated
using the projected Keplerian mask ($x,y,v_{\rm K}$) derived for the HD 163296 disk. 
To obtain the strictest upper limits on the integrated line intensity, 
we include only the positions and velocities associated with the disk.
\changetwo{Therefore, the mask cube contains pixels set equal to unity for ($x,y,v_{\rm K}$) positions
only, and all other pixels are set to zero.} The upper limit is set at 3$\sigma$ 
\change{where $\sigma = \delta v \sqrt{N} \sigma_{\mathrm{rms}}$,
$\delta v$ is the velocity channel width in km s$^{-1}$, $N$ is the number of independent measurements 
% at a given ($x,y,v_{\rm K}$) position that satisfy
contained within the projected Keplerian mask, and $\sigma_{\mathrm{rms}}$ 
is the rms noise per channel in mJy beam$^{-1}$ (see Table~\ref{tab:obs_par}).
To account for correlated noise within the size of the beam, 
we sum over all ($x,y,v_{\rm K}$) pixel positions
and divide by the number of pixels per beam $n_{\rm ppb}$, to get 
$N = \Sigma(x,y,v_{\rm K})/n_{\rm ppb}$, the number of 
independent measurements over the integrated Keplerian mask.}
The disk-integrated upper limits for each CH$_3$OH line are listed in
Table~\ref{tab:ch3oh_col_dens_abun}.

\subsubsection{Matched filter analysis in the $uv$ plane}
\label{sec:res_matchedfilter}

To maximize the SNR in the $uv$ plane, we apply a matched filter to the CH$_3$OH line
visibility data \citep{Loomis2018}. In this technique, a template image cube is 
sampled in $uv$ space to obtain a set of template visibilities that act as
the filter. The filter is then cross-correlated with a set of low SNR visibilities
(in this case, the CH$_3$OH data) in an attempt to detect any signal that is co-spatial 
with the template emission. \citet{Loomis2018} and \citet{Carney2017}
have published positive detections using the matched filter technique
\changethree{for CH$_3$OH and H$_2$CO}, which can 
provide an improvement in SNR of >50-500\% over the traditional aperture masking,
depending on the spectral resolution of the observed visibilities. 

We use the H$_2$CO 3$_{03}$--2$_{02}$ detection \changethree{towards the HD 163296 disk} reported in \citet{Carney2017} as the 
template emission profile under the assumption that CH$_3$OH and H$_2$CO 
\change{reside in similar regions. The emission morphologies will 
be dominated primarily by Keplerian rotation, therefore
a high degree of co-spatiality is expected.} 
The H$_2$CO line is re-imaged with CLEAN to achieve a 
spatial and spectral resolution equal to the observed
CH$_3$OH lines. Channels with H$_2$CO emission ($v = 1.6-10$ km s$^{-1}$) are
sampled in $uv$ space using the \textsc{python}
\texttt{vis\_sample}\footnote{\texttt{vis\_sample} is publicly 
available at \url{https://github.com/AstroChem/vis\_sample} or in the 
Anaconda Cloud at \url{https://anaconda.org/rloomis/vis\_sample}} routine.
The matched filter is run for the CH$_3$OH line visibility data individually
and after line stacking. 

Figure~\ref{fig:matchedfilter} shows the spectrum that is produced by the 
matched filter analysis for the band 7 CH$_3$OH data. The filter response 
in units of $\sigma$ is the measure of the SNR of the cross-correlation between the 
CH$_3$OH line visibility data and the filter derived from the template H$_2$CO emission. 
A correlation between the CH$_3$OH data and the filter would result in a peak at the source velocity. 
No such feature is seen in the filter response spectrum of any CH$_3$OH
lines in the HD 163296 disk, suggesting that the detection threshold for
methanol is well below the sensitivity achieved in our ALMA observations.
\changetwo{The matched filter analyses confirm the non-detection of CH$_3$OH found 
during analysis in the image plane.
The same analysis for the band 6 lines also results in no 
detection, which is expected given that the band 7 lines should be brighter.}

\subsection{CH$_3$OH column density and abundance upper limits}
\label{sec:res_abun_uplims}

\begin{table*}[!ht]
 \caption{Disk-averaged column density and abundance of CH$_3$OH in HD 163296 and TW Hya.}
 \centering
 \label{tab:ch3oh_col_dens_abun}
 \begin{tabular}{llccccrr}
 \hline \hline
 Object & Line & $\int I_{\nu} d\nu$\tablefootmark{$\dagger$} & $E_{\rm u}$ & log($A_{\rm ul}$)  & $n_{\rm crit}$\tablefootmark{a} & $N_{\rm avg}$ & ${\rm CH_3OH}/{\rm H_2}$ \\
 & & [mJy km s$^{-1}$] & [K] & [s$^{-1}$] & [cm$^{-3}$] & [cm$^{-2}$] & \\
 \hline \\
 HD 163296 & CH$_3$OH 5$_{05}$--4$_{04}$ (E) & $<51$ & 47.9 & $-4.22$ & 1.6(06) & $<6.9 (12)$ & $<2.1 (-11)$ \\ \\

  & CH$_3$OH 5$_{05}$--4$_{04}$ (A) & $<51$ & 34.8 & $-4.22$ & 4.3(05) & $<4.1 (12)$ & $<1.3 (-11)$ \\ \\

  & CH$_3$OH 1$_{10}$--1$_{01}$ (A) & $<26$ & 16.9 & $-3.49$ & 4.3(07) & $<7.0 (11)$ & $<2.2 (-12)$ \\ \\

  & CH$_3$OH 2$_{11}$--2$_{02}$ (A) & $<26$ & 21.6 & $-3.49$ & 5.0(06) & $<5.0 (11)$ & $<1.6 (-12)$ \\ \\
%   \hline \\
 TW Hya & CH$_3$OH stacked\tablefootmark{*} & 26.5 $\pm$ 2.7\tablefootmark{c} & 28.6 & $-3.49$ & 3.0(06) & $4.7 (12)$ & $1.1 (-12)$ \\ \\
 \hline
 \end{tabular}
 \tablefoot{The disk-averaged column density is calculated using Equation~\ref{eq:col_dens} with $T_{\rm ex}$ = 25 K. 
 The format $a$($b$) translates to $a \times 10^b$. Flux errors are dominated by systematic uncertainties, taken to be 10\%. \\
 \tablefoottext{$\dagger$}{Upper limits are derived  at the 3$\sigma$ level using the HD 163296 Keplerian mask (see Section~\ref{sec:res}).} \\
 \tablefoottext{*}{The stacked detection consists of three CH$_3$OH transitions: 
 CH$_3$OH 2$_{11}$--2$_{02}$ (A) at 304.208 GHz, 
 CH$_3$OH 3$_{12}$--3$_{03}$ (A) at 305.472 GHz, 
 and CH$_3$OH 4$_{13}$--4$_{04}$ (A) at 307.166 GHz. 
 Excitation parameters for the CH$_3$OH 3$_{12}$--3$_{03}$ (A) \changeref{line}
 are used to calculate column density.}\\
  \textbf{References:}  \tablefoottext{a}{\citet{Rabli2010};} \tablefoottext{b}{\citet{Walsh2016b}.}
 }
\end{table*}

We estimate the disk-averaged column density
of CH$_3$OH based on the integrated
line intensity upper limit, an assumed excitation 
temperature, and the total disk mass. Following the formula used by 
\citet{Remijan2003} and \citet{Miao1995} for optically thin emission
in local thermodynamic equilibrium (LTE), we can estimate the column density

\begin{equation}
 \label{eq:col_dens}
 N = 2.04 \frac{\int I_{\nu} dv}{\theta_{\rm a} \theta_{\rm b}} \frac{Q_{\rm rot} \ {\rm exp} (E_{\rm u} / T_{\rm ex})}{\nu^3 \langle S_{\rm ul} \mu^2 \rangle}
 \times 10^{20} \ {\rm cm}^{-2},
\end{equation}

\noindent where $\int I_{\nu} dv$ is the integrated line intensity in Jy beam$^{-1}$ km s$^{-1}$,
$\theta_{\rm a}$ and $\theta_{\rm b}$ correspond to the
semi-major and semi-minor axes of the synthesized beam in arcseconds,  
$T_{\rm ex}$ is the excitation temperature in K, and
$\nu$ is the rest frequency of the transition in GHz. 
The partition function ($Q_{\rm rot}$), upper energy level 
($E_{\rm u}$, in K), and the temperature-independent
transition strength and dipole moment ($S_{\rm ul} \mu^2$, in debye$^2$)
for CH$_3$OH are taken from the \changeref{Cologne Database for Molecular Spectroscopy
\citep[CDMS;][]{Muller2005}.}

Methanol is expected to form primarily in ice \changethree{in cold regions}
of protoplanetary disks, where gas densities are higher \changeref{\citep[$\sim$10$^9$ cm$^{-3}$;][]{Walsh2014a}}
than the critical density of the observed CH$_3$OH transitions
\citep[$10^6 - 10^7$ cm$^{-3}$;][]{Rabli2010}. Recent physical models
of the HD 163296 disk have gas densities >$10^6$ cm$^{-3}$ in the region 
$z/r < 0.4$ \citep{Qi2011,deGregorioMonsalvo2013,Rosenfeld2013}, where $z$ and $r$ are the disk 
height and radius, respectively. In recent models of the TW Hya disk,
\citet{Walsh2016b} varied the methanol emitting region \changetwo{over the range
$z/r < 0.1$, $0.1 < z/r < 0.2$, and $0.2 < z/r < 0.3$, which all fit the data equally well. 
These models all had methanol present at $z/r < 0.3$, suggesting that emission is arising from
dense regions within the disk.} Under these conditions, LTE is a reasonable assumption,
\changetwo{and thus $T_{\rm ex}$ \changethree{is expected to} equal the kinetic temperature of the gas.}

% We explore excitation temperatures of 25, 50, and 75 K.
Assuming optically thin emission, the disk-averaged column density can be
used to estimate the total number of CH$_3$OH molecules
in the disk $N({\rm CH_3OH}) = N_{\rm avg}(a \times b)$,
where $(a \times b)$ is the total emitting area of the disk.
Assuming the total disk mass is primarily molecular hydrogen,
we can estimate the total number of H$_2$ molecules
$N({\rm H_2}) = M_{\rm disk} / m_{\rm H_2}$, where $m_{\rm H_2}$
is the molecular hydrogen mass.
The CH$_3$OH emitting area is set to $a = b = 7\arcsec$ based on the
H$_2$CO emission diameter in \changeref{the HD 163296 disk} \citep{Carney2017}, assuming a 
similar chemical origin and distribution. The total
disk mass is $\sim$$0.09 \ M_{\odot}$ \changethree{based on models 
of CO observations} \citep{Qi2011,Rosenfeld2013}.
Table~\ref{tab:ch3oh_col_dens_abun} shows the disk-averaged column 
density and abundance for \changethree{the single temperature assumption} 
$T_{\rm ex}$ = 25 K in LTE, which is approximately the same as the excitation temperature found for 
H$_2$CO in the HD 163296 disk \citep{Qi2013,Carney2017}. %, 50, and 75 K. 
The CH$_3$OH 2$_{11}$--2$_{02}$ (A) line provides the strictest upper
limit on the methanol \changethree{column density and abundance in HD 163296, with 
$N_{\rm avg} \lesssim 5.0 \times 10^{11}$ \changeref{cm$^{-2}$} and ${\rm CH_3OH}/{\rm H_2} \lesssim 1.6 \times 10^{-12}$,}
based on its disk-integrated line intensity upper limit
\changetwo{and assuming an excitation temperature of $T_{\rm ex} = 25$ K.}
Table~\ref{tab:app_ch3oh_col_dens_abun} in the Appendix shows the disk-averaged column 
density and abundance for a range of LTE excitation conditions with
$T_{\rm ex}$ = 25, 50, and 75 K. The abundances do not vary with 
$T_{\rm ex}$ by more than a factor of 2--3 in the most extreme cases.
% with $T_{\rm ex} = 25$ K in LTE for excitation conditions, 
% which is representative of the disk midplane.

\subsection{H$_2$CO and CH$_3$OH in HD 163296 and TW Hya}
\label{sec:res_h2co_ch3oh}

\begin{table*}[!tp]
 \caption{Disk-averaged column density and abundance of H$_2$CO in HD 163296 and TW Hya.}
 \centering
 \label{tab:h2co_col_dens_abun}
%  \resizebox{\textwidth}{!}{
 \begin{tabular}{llccccccc}
 \hline \hline
 Object & Line & $\int I_{\nu} d\nu$ & $E_{\rm u}$ & log($A_{\rm ul}$) & $n_{\rm crit}$\tablefootmark{a} & $N_{\rm avg}$ & ${\rm H_2CO}/{\rm H_2}$ & ${\rm CH_3OH}/{\rm H_2CO}$\tablefootmark{$\dagger$} \\
 & & [mJy km s$^{-1}$] & [K] & [s$^{-1}$] & [cm$^{-3}$] & [cm$^{-2}$] &  & \\
 \hline \\
 HD 163296 & H$_2$CO 3$_{12}$--2$_{11}$ & 890 $\pm$ 89\tablefootmark{b} & 33.4 & $-3.55$ & 5.7(06) & 2.1(12) & 6.3($-$12) & < 0.24 \\ \\
 TW Hya & H$_2$CO 3$_{12}$--2$_{11}$ & 291 $\pm$ 29\tablefootmark{c} & 33.4 & $-3.55$ & 5.7(06) & 3.7(12) & 8.9($-$13) & 1.27 $\pm$ 0.13  \\ \\
 \hline
 \end{tabular}
%  }
 \tablefoot{The disk-averaged column density is calculated using Equation~\ref{eq:col_dens} with $T_{\rm ex}$ = 25 K. 
 The format $a$($b$) translates to $a \times 10^b$. Flux errors are dominated by systematic uncertainties, taken to be 10\%. \\
 \tablefoottext{$\dagger$}{Ratios are determined using the CH$_3$OH disk-integrated column 
 density from Table~\ref{tab:ch3oh_col_dens_abun}. HD 163296: based on 
 the strictest upper limit from the CH$_3$OH 2$_{11}$--2$_{02}$ (A) line.} 
 TW Hya: based on the stacked CH$_3$OH detection.\\
 \textbf{References:} \tablefoottext{a}{\citet{Wiesenfeld2013};} \tablefoottext{b}{\citet{Qi2013};} \tablefoottext{c}{\citet{Oberg2017}.}
 }
\end{table*}

We estimate the fraction of methanol relative to formaldehyde based on our
upper limits for CH$_3$OH in HD 163296 and compare to the TW Hya disk,
the only \changethree{Class II} protoplanetary disk for which there is a gas-phase methanol detection \citep{Walsh2016b}.
Integrated line intensities for H$_2$CO detections in HD 163296 and TW Hya are taken
from the literature, and their disk-averaged column densities and abundances are derived in the 
same manner as described in Section~\ref{sec:res_abun_uplims} to ensure consistency
when comparing the H$_2$CO and CH$_3$OH content.
The TW Hya disk mass is $0.05 \ M_{\odot}$ based on observations of the HD molecule \citep{Bergin2013}.
The emitting area for H$_2$CO in TW Hya is set to $a = b = 3\arcsec$ based on the 
diameter of emission observed by \citet{Oberg2017}. \changetwo{The same $3\arcsec$ emitting
area is used for CH$_3$OH in TW Hya.} 
Table~\ref{tab:h2co_col_dens_abun} shows
the calculated column densities and abundances for the H$_2$CO observations.

For HD 163296, the CH$_3$OH 2$_{11}$--2$_{02}$ (A) line 
is used to calculate the methanol-to-formaldehyde ratio as it gives the strictest
upper limits on the methanol abundance. For TW Hya, we obtained the integrated
line intensity of the stacked methanol detection by \citet{Walsh2016b}, assume that
the majority of emission is due to the strongest individual line \citep[CH$_3$OH 3$_{12}$--3$_{03}$ (A)
at 305.473 GHz \changethree{with $E_{\rm u}$ = 28.6 K}:][]{Walsh2014a,Loomis2018}, and use the excitation parameters of that 
line with Equation~\ref{eq:col_dens} to derive the TW Hya CH$_3$OH 
column density and abundance, and subsequently the ${\rm CH_3OH}/{\rm H_2CO}$ ratio
for the disk.

Results for the ${\rm CH_3OH}/{\rm H_2CO}$ \changeref{ratio} in TW Hya and HD 163296 
can be found in Table~\ref{tab:h2co_col_dens_abun}. 
% Ratios calculated with the H$_2$CO 3$_{12}$--2$_{11}$ line are 
% more representative of the true CH$_3$OH/H$_2$CO ratio since
% the H$_2$CO 3$_{12}$--2$_{11}$ upper energy level ($E_{\rm u}$) and Einstein A coefficient ($A_{\rm ul}$)
% are closer to that of the band 7 methanol lines observed in these disks.
Ratios calculated with the H$_2$CO 3$_{12}$--2$_{11}$ line should
be representative of the true CH$_3$OH/H$_2$CO ratio since
the H$_2$CO 3$_{12}$--2$_{11}$ upper energy level ($E_{\rm u}$), Einstein A coefficient ($A_{\rm ul}$),
and critical density ($n_{\rm crit}$) are similar to that of the band 7 methanol lines observed in these disks.
\changethree{Thus, we obtain CH$_3$OH/H$_2$CO ratios of < 0.24 for HD 163296 and
1.27 for TW Hya,} which suggests that the disk around HD 163296 is
less abundant in methanol relative to formaldehyde \changeref{compared to the TW Hya disk.}

% The disk-integrated line intensity metric results in 
% a lower estimates for the total number of molecules and for the 
% CH$_3$OH/H$_2$CO ratio in both disks compared to the
% model-based ratio derived from model column densities. In both cases, the 
% upper limit on the CH$_3$OH abundance in the HD 163296 disk is less than the
% abundance of CH$_3$OH from the detection in TW Hya.

% In addition to the disk-integrated line intensity (or upper limit), we also
% use recent models of H$_2$CO and CH$_3$OH in HD 163296 and TW Hya 
% for a secondary comparison of the molecular content between these disks.
% TW Hya column density profiles is extracted for H$_2$CO based on the 
% best-fit ``flared'' model from \citet{Oberg2017} and for CH$_3$OH based on the 
% best-fit low $z/r$ model from \citet{Walsh2016b}. The  
% disk-integrated column density is used to determine the total number of molecules 
% in the disk following the method in Section~\ref{sec:res_abun_uplims}. From the 
% total number of each molecular species we
% obtain a model-based methanol-to-formaldehyde ratio for TW Hya.
% To determine this ratio in the HD 163296 disk 
% we use the best-fit H$_2$CO model from \citet{Carney2017}, which is based
% on a physical model by \citet{Qi2011}, and
% scale the CH$_3$OH abundance by a constant factor. The
% H$_2$CO model has an abundance structure $X_{\rm in}/X_{\rm out} = 0.5$
% where $X_{\rm in}$ = $4.0\times 10^{-12}$ relative to H$_2$ for $r < 270$ AU.

\subsection{Model CH$_3$OH spectra for HD 163296}
\label{sec:res_model_ch3oh}

\changethree{In addition to the extraction methods described \changeref{in previous sections}, we also 
attempt a forward modeling approach to interpret the CH$_3$OH non-detections 
toward HD 163296.} We model the HD 163296 CH$_3$OH band 7 spectra \changetwo{using 
\change{a parameterized disk structure} and radiative 
transfer methods} in order to compare the modeled emission
to the noise level in the 
Keplerian-masked image cubes. We adopt the physical structure and the 
abundance structure of the model used by \citet{Carney2017} to reproduce ALMA
observations of H$_2$CO in the HD 163296 disk,
then scale the CH$_3$OH abundance with respect to the H$_2$CO abundance.
The \changeref{Line Modeling Engine \citep[LIME;][]{Brinch2010}} 3D radiative transfer code is run in LTE with 
10000 grid points at the source distance of the original 
\citet{Qi2011} physical model ($d=122$ pc) to create
synthetic images of the CH$_3$OH observations. The synthetic images are 
continuum-subtracted, sampled in $uv$ space with the \textsc{python}
\texttt{vis\_sample} routine, and imaged with CLEAN at the same velocity resolution
as the observations.

\begin{figure}[!tp]
 \centering
 \includegraphics[width=0.4\textwidth]{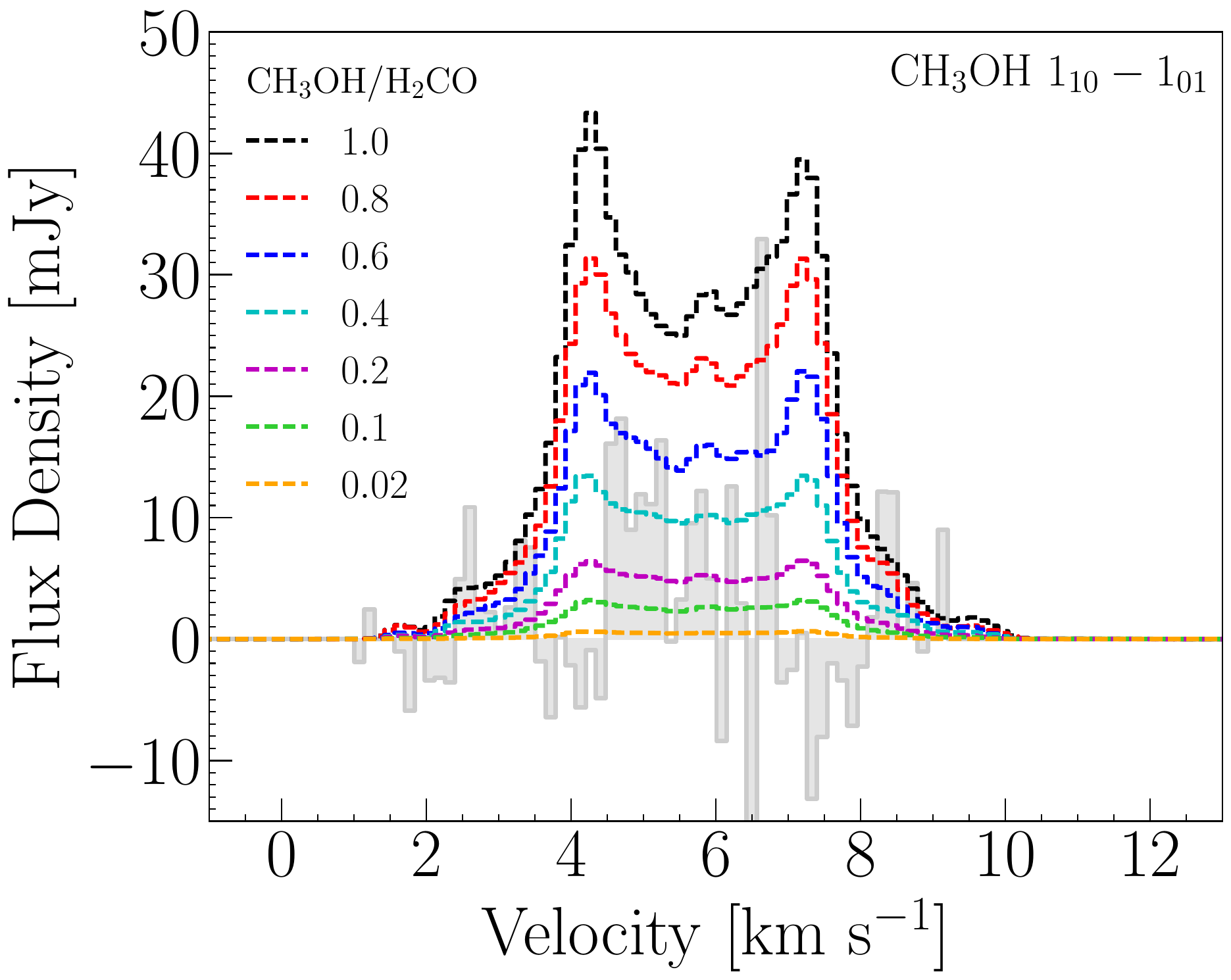} \\
 \vspace{0.3cm}
 \includegraphics[width=0.4\textwidth]{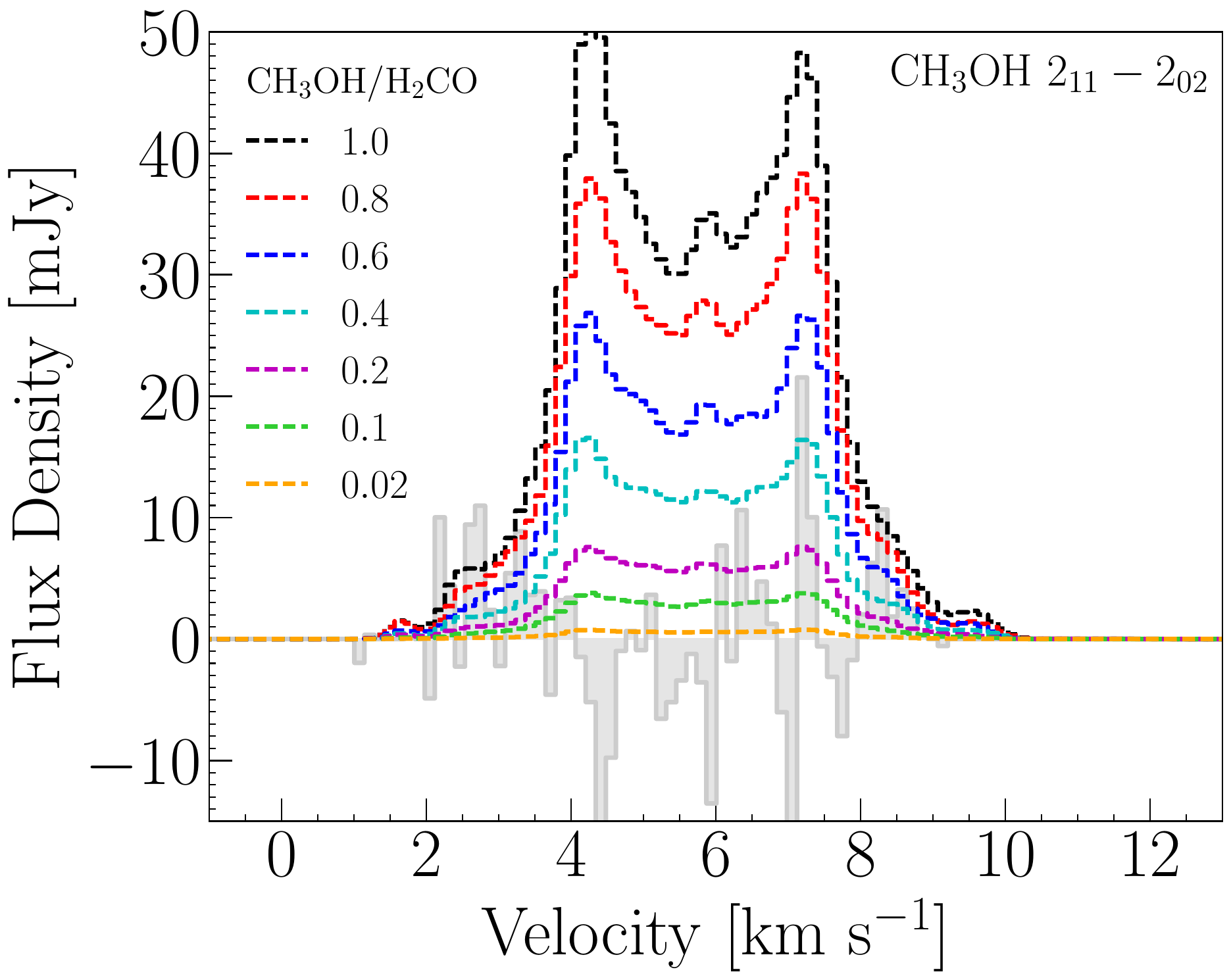} \\
 \vspace{0.3cm}
 \includegraphics[width=0.4\textwidth]{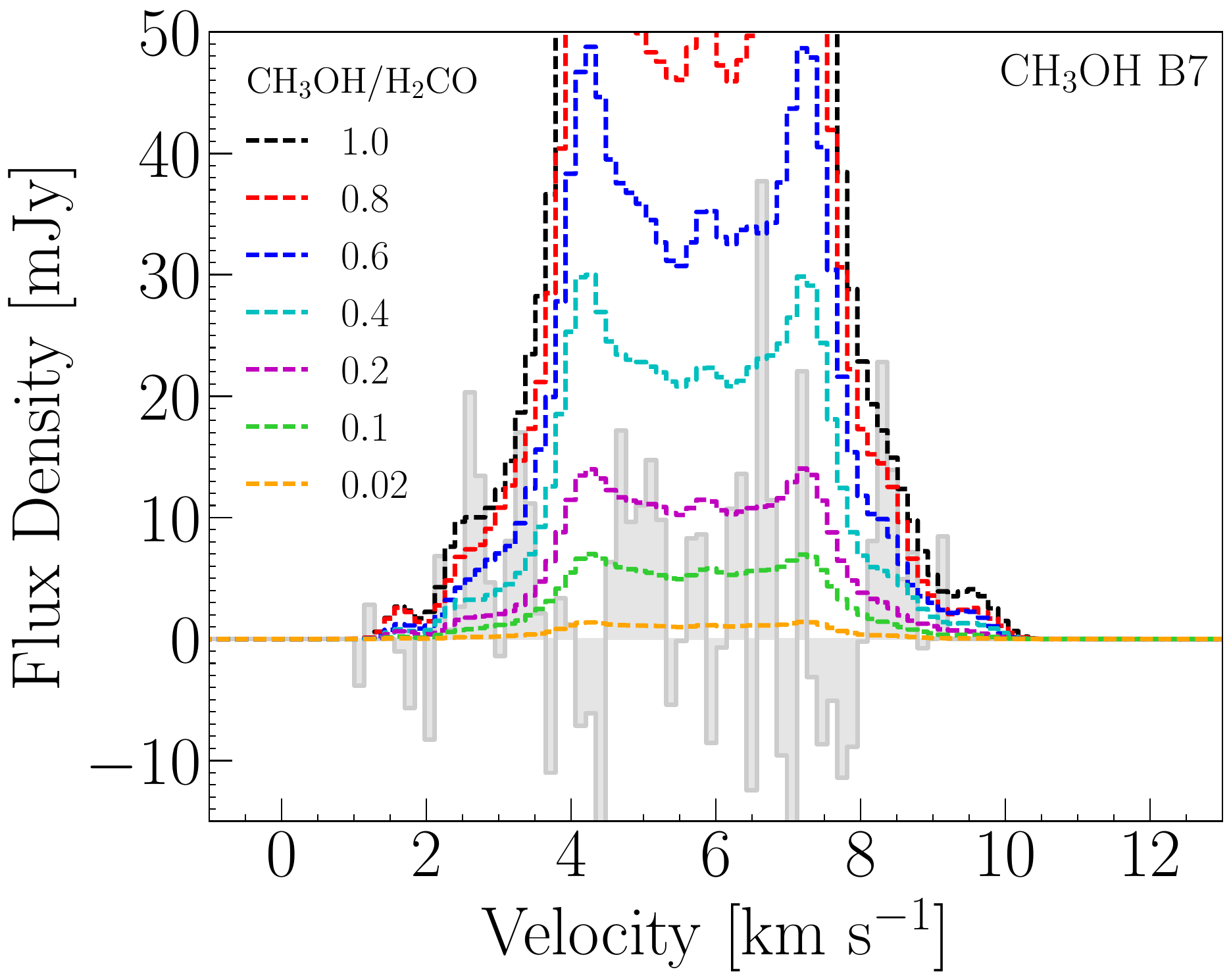}
 \caption{Model \changeref{CH$_3$OH} spectra at different CH$_3$OH/H$_2$CO abundance ratios (colored dashed lines) compared to the \changeref{ALMA} CH$_3$OH non-detections 
 after Keplerian masking (gray) in the HD 163296 disk. Given the \changeref{sensitivity levels achieved, the ALMA observations} should be sensitive
 to the presence of methanol in the disk for CH$_3$OH/H$_2$CO $\gtrsim 0.2$ based on the stacked band 7 lines.
 (Top) CH$_3$OH 1$_{10}$--1$_{01}$ (A) line. (Middle) CH$_3$OH 2$_{11}$--2$_{02}$ (A) line. 
 (Bottom) Stacked band 7 CH$_3$OH lines. 
 }
 \label{fig:mod_spectra}
\end{figure}

Figure~\ref{fig:mod_spectra} shows the disk-integrated \changeref{model} spectra for the CH$_3$OH 
band 7 lines for a range of methanol-to-formaldehyde ratios, as indicated by the legend.
The spectra show that a line should have been detected \changethree{in the disk
around HD 163296} for a 
CH$_3$OH/H$_2$CO ratio of $\sim$0.2 for the most sensitive case (stacked band 7 lines).
\changethree{This result is consistent with the upper limit on this ratio 
derived from the integrated intensity of the Keplerian mask cube 
as presented in Section~\ref{sec:res_h2co_ch3oh}.}

\section{Discussion}
\label{sec:disc}

The results presented in Table~\ref{tab:h2co_col_dens_abun} suggest that 
the HD 163296 disk has a lower overall gas-phase methanol content with respect to
formaldehyde than the TW Hya disk. In this section we discuss possible reasons
for a lower CH$_3$OH/H$_2$CO ratio in HD 163296, as well as a brief assessment of
the observing time needed to detect the low predicted abundances of gas-phase methanol in this disk.

\subsection{The CH$_3$OH/H$_2$CO ratio in HD 163296 and TW Hya}

It should be noted that there are uncertainties on the order of a 
factor of a few when deriving the CH$_3$OH/H$_2$CO ratio as described in this work. Namely, the 
\changeref{column density} calculation for the methanol detection in TW Hya is a result of three stacked line transitions
rather than a single common transition as for H$_2$CO observed in both disks.
The CH$_3$OH 3$_{12}$--3$_{03}$ (A) line at 305.473 GHz is the strongest methanol line observed
in TW Hya, but it is not the sole contributor to the detected line emission.
However, even if \changethree{all three lines are equally strong and the 305.473 GHz line contributes 
only 33\% to the total stacked line intensity, then the inferred TW Hya CH$_3$OH/H$_2$CO ratio of 
0.42 is still higher than our upper limit for HD 163296 of < 0.24.
Matched filter analysis of the TW Hya CH$_3$OH detections \citep[e.g., Figure 7 in][]{Loomis2018}
shows that the CH$_3$OH 3$_{12}$--3$_{03}$ (A) line is indeed
stronger than the other two band 7 lines used by \citet{Walsh2016b} for line stacking, suggesting 
that a contribution of $\sim$50\% to the stacked emission is a reasonable estimate.}

Modeling by \citet{Willacy2007} explored complex gas-grain chemical models of protoplanetary
disks including H$_2$CO and CH$_3$OH with the following desorption processes: 
thermal desorption, desorption due to cosmic-ray heating of grains, and photodesorption. 
Their models, \changethree{based on the UMIST Database for Astrochemistry network}, show that outer 
disk abundances should give CH$_3$OH/H$_2$CO $\approx$ 0.04, \changeref{which is lower than both the ratio} 
found for \changethree{TW Hya and \changeref{the upper limit on the ratio found for} HD 163296.}
However, \change{these models} neglected radical-radical pathways to form larger complex organic molecules.
\changethree{Gas-grain chemical models by \citet{Semenov2011} based on the Ohio State University 
(OSU) network predict low column densities of methanol ice due to the high diffusion 
barrier used in the grain-surface chemistry, which highlights the importance of the assumed 
chemical parameters in these models. Their models and work by \citet{Furuya2014} show
that production of CH$_3$OH is sensitive to turbulent mixing and that the abundance of gas-phase 
CH$_3$OH, and thus the CH$_3$OH/H$_2$CO ratio, will increase when turbulent mixing is strong.
The HD 163296 disk has a low degree of turbulence $\lesssim$0.05 $c_{\rm s}$ 
\citep{Flaherty2015,Flaherty2017}, while the TW Hya disk has similar low values of $\lesssim$0.05--0.10 $c_{\rm s}$ \citep{Flaherty2018},
suggesting vertical mixing is not strong in these disks.}

Other recent work by \citet{Walsh2014a} based on the OSU network investigates the production
of complex molecules in disks, including H$_2$CO and CH$_3$OH, using an extensive full chemical network 
with chemical ingredients similar to the previously mentioned works. 
\changeref{Their models include two-body, X-ray, and cosmic ray reactions and
photoreactions in the gas phase and on grain surfaces as all as gas-grain reactions (e.g., freeze-out
and photodesorption) around a T Tauri-like PMS star.}
% The \citet{Walsh2014a} models include gas-phase 
% two-body reactions, photoreactions, cosmic-ray and X-ray reactions, gas-grain interactions, 
% grain-surface two-body reactions, grain-surface photoreactions, and grain-surface cosmic-ray-induced 
% and X-ray induced photoreactions. 
The models in that work show that \changeref{their} outer disk ($R=250$ AU)
hosts a large methanol and formaldehyde ice reservoir with a sufficient 
number of these molecules released into the gas phase to
% have column densities of $N = 1.7\times10^{12}$ cm$^{-2}$ for H$_2$CO and $N = 7.2\times10^{12}$ cm$^{-2}$
% for CH$_3$OH at $R=250$ AU. These column densities yield an outer disk methanol-to-formaldehyde ratio of
give CH$_3$OH/H$_2$CO $\approx$ 0.33. \changeref{Subsequent work by \citet{Walsh2015} examines
molecular complexity across different luminosities (M dwarf, T Tauri, Herbig Ae/Be) for the inner disk
following a similar modeling approach. The authors find that molecular organics like H$_2$CO and 
CH$_3$OH contribute to the disk gas-phase carbon and oxygen reservoir for the cooler PMS stars, but
not for the warmer Herbig Ae/Be PMS stars. These modeling results}
% \changethree{which is less than the value found for the TW Hya disk, 
% but similar to what is estimated for the HD 163296 based on the upper limits presented here.} 
% Their model is based on a \changeref{disk around a} typical T Tauri star \changethree{but 
% does not reproduce the TW Hya CH$_3$OH/H$_2$CO ratio well,} 
perhaps point to important differences in how these two molecules are formed in 
T Tauri disks versus Herbig Ae/Be disks.

% In the recent paper by Agundez et al. (2018) comparing chemistry between T Tauri and 
% Herbig Ae disks, it is stated that models that neglect grain-surface chemistry can reproduce 
% the observed column density of gas-phase H2CO towards T Tauri and Herbig Ae stars (~ 10^{12} cm-2).

The underlying physical structure in the TW Hya disk and the HD 163296 disk might explain the observed
discrepancy between their CH$_3$OH/H$_2$CO ratios. Recent observations of submillimeter and scattered 
light in these disks highlight important differences in their dust structure. 
The micron-sized dust observed in scattered light is highly coupled to the gas and traces 
the surface layers of the disk, while millimeter-sized dust has mostly decoupled from the gas
and settled toward the disk midplane \citep{Dullemond2004,DAlessio2006,Williams2011}. The TW Hya disk
was observed with ALMA in the band 6 continuum at 850 $\mu$m and with VLT/SPHERE in $H-$band at 
1.62 $\mu$m \citep{Andrews2016,vanBoekel2017}, showing several rings and gaps in both millimeter- 
and micron-sized dust. The micron-sized dust rings tracing the surface layers 
extend beyond the millimeter-sized dust in this disk.
In contrast, recent scattered light observations by VLT/SPHERE in $H-$band, Keck/NIRC2 in $J-$band, 
and ALMA 1.3 millimeter observations of the HD 163296 disk show that no scattered light is observed
beyond the innermost millimeter dust ring, suggesting that the surface layers of the 
outer disk are relatively flat and may be shadowed by the innermost dust ring \citep{Muro-Arena2018,Guidi2018}.
Ultraviolet radiation from the central star can release molecular ices
back into the gas phase via UV photodesorption \citep{Oberg2009,Oberg2015,Huang2016},
% or thermal desorption due to a temperature inversion if dust is heated by (inter)stellar radiation 
% in the outer disk \citep{Cleeves2016}.
\changetwo{which may be suppressed if the HD 163296 outer disk is shadowed.}
% If the HD 163296 outer disk is shadowed and remains partially shielded to UV photons, 
% \change{it may be that only UV photodesorption is operating there and thus releasing
% H$_2$CO into the gas phase efficiently while CH$_3$OH is readily destroyed
% upon release \citep{Oberg2009,CruzDiaz2016}. On the other hand, TW Hya may 
% have an outer disk warm enough for thermal desorption to operate, which 
% can be a better mechanism to produce equal amounts of gas-phase H$_2$CO and CH$_3$OH at 
% low temperatures, e.g., due to co-desorption with CO \citep{Ligterink2018}. 
% The HD 163296 disk has already shown hints of a temperature inversion at large radii as inferred from 
% dust evolution models \citep{Facchini2017} and as traced by DCO$^+$ \citep{Salinas2018},
% but this inversion may not be enough to produce a detectable amount of gas-phase CH$_3$OH.}

\changetwo{Alternatively, both disks may have a similar degree of UV irradiation, but 
as a Herbig Ae star HD 163296 will have a harder UV spectrum than TW Hya, which 
is dominated by Lyman-$\alpha$ emission \citep[e.g., Figure 1 in][]{Walsh2015}.
The UV photodesorption rate of methanol ice is a strong function of photon energy and 
absorption cross section \citep{CruzDiaz2016}, and therefore will depend on the shape of the radiation field as well 
as the strength \citep{Bertin2016}. A harder, stronger Herbig Ae radiation field will lead to more 
CH$_3$OH fragmentation upon photodesorption and thus methanol ice will be converted
into other gas-phase species which could go on to seed H$_2$CO formation in the gas phase.}

\changetwo{Another possibility is that the HD 163296 disk formed from a protostar that 
did not inherit a large \changeref{amount} of methanol ice. Perhaps during formation, temperatures
remained too warm for CO freeze-out needed to produce the high CH$_3$OH/H$_2$O ice ratios seen in ISM ices. 
Chemical models with some methanol already formed at earlier stages \citep{Walsh2014a} host a 
more abundant methanol ice reservoir than models which start from atomic abundances, which have
orders of magnitude lower methanol ice abundances \citep[e.g.,][]{Molyarova2017}.}
% The photodesorption processes of H$_2$CO and CH$_3$OH must be thoroughly checked with detailed models, which is 
% beyond the scope of this work.}

% Another possibility is that methanol is simply not formed efficiently on grains in HD 163296.
% Methanol can be a gauge of the availability of atomic hydrogen in disks.
% The formation of CH$_3$OH requires sufficiently abundant atomic hydrogen to 
% overcome the reaction barriers for H$_2$CO + H and subsequently CH$_2$OH + H or CH$_3$O + H.
% The flux of H atoms on CO ice has been shown to influence the hydrogenation
% process and thus the formation of methanol \citep{Watanabe2003,Fuchs2009}.
% Less methanol in the HD 163296 disk could indicate less atomic hydrogen available to bombard 
% the icy grains in the disk midplane.

While both formaldehyde and methanol are thought to be formed via hydrogenation of CO ices
\citep{Watanabe2002}, formaldehyde can also be formed in the \changeref{gas phase}. 
\changethree{Recent chemical models by \citet{Agundez2018} that do not include grain-surface chemistry 
are able to reproduce observed column densities of H$_2$CO, but not CH$_3$OH, in the outer regions of T Tauri and Herbig Ae/Be disks.}
\changetwo{Reactions between CH$_3$ and atomic oxygen can occur in the disk surface layers where 
oxygen-bearing species are photodissociated \citep{Fockenberg2002,Atkinson2006}. This reaction,
however, has not been shown to contribute significantly to the H$_2$CO abundance in recent chemical models 
of disks around T Tauri stars \citep{Walsh2014a}. The contribution may be larger in warmer, 
strongly irradiated disks around Herbig Ae/Be stars.
Ion-molecule chemistry -- which has a large influence on the gas-phase reservoir in the 
intermediate layers of protoplanetary disks -- involving e.g., HCO$^+$,
H$_3$O$^+$, and H$_3^+$ may also contribute to the overall gas-phase H$_2$CO abundance \citep{Vasyunin2008}.}
\change{It may be that the HD 163296 disk is particularly rich in H$_2$CO formed in the \changeref{gas phase},
thus reducing its overall CH$_3$OH/H$_2$CO ratio.} 
\changeref{However, results from a recent analysis using the ortho-to-para ratio of H$_2$CO as a tool to investigate 
its chemical origins are consistent with significant grain-surface formation and subsequent desorption \citep{Guzman2018}.}
Detailed chemical models of the HD 163296 
protoplanetary disk beyond the scope of this work are required to test and quantify the 
importance of the production and destruction routes for H$_2$CO and CH$_3$OH discussed here.

\subsection{Detectability of methanol}

% ALMA sensitivity S = (k*Tsys)(A*N2(Np*bandwidth/channel width * time)^1/2)^-1 --> time

We can estimate the required ALMA observing time for a 3$\sigma$ detection of CH$_3$OH \changeref{in the HD 163296 disk} given 
a range of CH$_3$OH/H$_2$CO ratios consistent with our \changethree{upper limit of $< 0.24$.}
% To obtain a 3$\sigma$ detection of the CH$_3$OH 2$_{11}$--2$_{02}$ (A) line in the HD 163296 disk, 
% we can easily estimate the required observing time with ALMA using the same observing parameters
% described in this work: 45 antennas, $\sim$1$\arcsec$ beam, and $\sim$0.15 km s$^{-1}$ spectral resolution.
We consider methanol abundances relative to formaldehyde of \changethree{0.20, 0.10, 0.05}, as these 
would be below our current 3$\sigma$ upper limit of $< 0.24$ listed in Table~\ref{tab:h2co_col_dens_abun}. 
To observe \changetwo{the CH$_3$OH 2$_{11}$--2$_{02}$ (A) line of methanol with similar spatial and spectral resolution}
at these assumed CH$_3$OH/H$_2$CO ratios, we would need to increase our 
sensitivity by factors of about \changethree{$\sim$1.5, 2.5, and 5,} respectively. Because the telescope sensitivity is inversely proportional
to the square root of the observing time, $\sigma_{\rm S} \propto 1/\sqrt{t}$, the time required to realize 
these increases in sensitivity would multiply
by factors of \changethree{2.25, 6.25, and 25,} respectively. Based on the band 7 observations presented here with 
105 minutes of total on-source time, these factors translate to total on-source times of 
\changethree{$\sim$4 hrs, $\sim$11 hrs, $\sim$44 hrs for methanol at 20\%, 
10\%, and 5\%} of the formaldehyde content in \changeref{HD 163296}, respectively.
The detection of \changethree{10\%} methanol relative to formaldehyde is a 
clear practical limit for the HD 163296 disk based on these required integration times.

% The inclination of a protoplanetary disk can have an effect on the observed line strength, but
% we find the effect of disk inclination on the detectability of CH$_3$OH to 
% be negligible. LIME models of the HD 163296 disk as described in Section~\ref{sec:res_h2co_ch3oh}
% are rerun for disk inclinations from $i = 0^{\circ}-90^{\circ}$ in steps of $5^{\circ}$ for
% a representative methanol-to-formaldehyde abundance ratio of CH$_3$OH/H$_2$CO = 0.10. 
% The disk-integrated line intensity for the band 7 CH$_3$OH lines varies by 
% only 10-30\% over the full range of inclinations.

Disk size has a significant effect on methanol detectability. Using our HD 163296 
model, we decrease the outer radius of the disk and scale the disk physical 
structure \changeref{(i.e., gas density and temperature)} proportionally
to test the effect of disk size on the band 7 methanol line strengths for Herbig disks similar to HD 163296. 
The LIME models are rerun for an outer disk radius from $R_{\rm out} = 100 - 600$ AU in steps of 50 AU for 
CH$_3$OH/H$_2$CO = 0.10. The disk-integrated line intensity for the band 7 CH$_3$OH lines 
decreases by one order of magnitude for disks with $R_{\rm out}$ = 250 AU and 
by more than two orders of magnitude for disks with $R_{\rm out}$ = 100 AU. It is
highly unlikely that methanol will be detected within an observing time of $<20$ hours 
in most disks smaller than $\sim$300 AU, considering the difficulty in detecting methanol relative to formaldehyde
at the \changethree{$<25\%$ level} in the HD 163296 disk, which has a \changeref{radius of} $\sim$550 AU
and a proximity closer than most nearby star-forming regions.
\changetwo{These results depend on the assumption that CH$_3$OH 
shares the same extended emitting area as H$_2$CO.}
% If the CH$_3$OH emitting region is compact 
% relative to the size of the disk, the influence of overall disk size on the 
% detectability of methanol will be suppressed. }
% What is more important is the column density within the emitting area.}

\changetwo{It may be that the methanol lines targeted in this work are not suitable candidates 
for disks around Herbig Ae/Be stars. The choice to target these four CH$_3$OH lines 
with ALMA in band 6 and band 7 was motivated by 
the chemical modeling of a disk around a T Tauri star \citep{Walsh2014a} and by the methanol detection in 
the disk around TW Hya, also a T Tauri star \citep{Walsh2016b}. Disks around Herbig Ae/Be stars are 
warmer, with a larger thermally desorbed inner reservoir due to the stronger stellar radiation. 
There is a potential reservoir of hot methanol in the inner disk atmosphere, similar to the hot 
water reservoir already observed in disks around less luminous T Tauri stars \citep{Carr2008,Salyk2008}.
Such emission could be compact yet still accessible in Herbig Ae/Be disks.}

\change{In summary, the CH$_3$OH lines in ALMA band 7 presented here should be 
detectable in disks with a CH$_3$OH/H$_2$CO ratio down to \changethree{$\sim$10\%} within realistic 
observing times, but only in disks with similar mass, size, distance, and H$_2$CO abundance 
as those found in the HD 163296 disk.}

\section{Conclusions}
\label{sec:concl}

This paper presents ALMA observations \changeref{targeting} two CH$_3$OH lines in band 6 and two
CH$_3$OH lines in band 7 in the
protoplanetary disk around HD 163296. We determine upper limits on the abundance of methanol
likely to be present in the HD 163296 disk and compare to TW Hya,
currently the only \change{Class II} disk with a positive detection of \change{gas-phase} methanol.
The conclusions of this work are as follows:

\begin{itemize}
  \item None of the four CH$_3$OH lines are detected in the disk around HD 163296
  individually nor after line stacking. \change{Upper limits on the integrated intensity at the 3$\sigma$ level 
  are < 51 mJy km s$^{-1}$ for band 6 lines and < 26 mJy km s$^{-1}$ for band 7 lines.}
  Neither aperture masking in the image plane, Keplerian masking in the image plane,
  nor matched filter analysis in the $uv$ plane recover any methanol emission, indicating
  that our calculated 3$\sigma$ upper limits are highly robust.
 
 \item The CH$_3$OH 2$_{11}$--2$_{02}$ (A) line provides the strictest upper limit on the 
 disk-averaged \changethree{column density and} abundance of methanol in the HD 163296 disk, \changethree{with 
 $N_{\rm avg} < 5.0 \times 10^{11}$ cm$^{-2}$ and ${\rm CH_3OH}/{\rm H_2}$ $\lesssim 1.6 \times 10^{-12}$
 at the 3$\sigma$ level.}
 
 \item The upper limit on the methanol-to-formaldehyde ratio in the HD 163296 disk is 
 \changethree{CH$_3$OH/H$_2$CO $<0.24$} at the 3$\sigma$ level. This ratio is lower than that of the 
 TW Hya disk at \changethree{CH$_3$OH/H$_2$CO = 1.27 $\pm$ 0.13,} 
%  and consistent with} recent model predictions at CH$_3$OH/H$_2$CO $\approx$ 0.33, 
 indicating that \changethree{the HD 163296 disk has a low amount
 methanol with respect to formaldehyde relative to the TW Hya disk.}
 
 \item Possible explanations for the lower CH$_3$OH/H$_2$CO ratio in HD 163296 include:
 a low amount of gas-phase methanol is desorbed from icy grains at the disk midplane
 due to the flatter, shadowed disk geometry as seen in recent images taken by VLT/SPHERE;
%  methanol formation on grains might be less efficient in HD 163296 if 
%  the atomic hydrogen flux near the disk midplane is significantly lower than in the TW Hya disk;
 \change{differences in the desorption processes in the HD 163296 disk compared 
 to the TW Hya disk;
 and a higher-than-expected gas-phase formaldehyde abundance, as H$_2$CO may also 
 be formed in the \changeref{gas phase} \changethree{in the disk upper layers}.}
 
 \item To detect methanol at the 3$\sigma$ level in the HD 163296 disk, we estimate that it is necessary
 to increase the total on-source observing time with the full ALMA \changeref{12-meter} array up 
 to \changethree{4 hours} to be sensitive to CH$_3$OH/H$_2$CO $\approx$ \changethree{20\%}
 and up to \changethree{11 hours} to be sensitive to CH$_3$OH/H$_2$CO 
 $\approx$ \changethree{10\%.} \change{These estimates apply to other Herbig Ae/Be disks
 with masses, sizes, and distances similar to that found for the HD 163296 disk.}

\end{itemize}

\begin{acknowledgements}
      The authors acknowledge support by Allegro, the European
      ALMA Regional Center node in The Netherlands, and expert
      advice from Luke Maud.
      M.T.C. and M.R.H. acknowledge support from the
      Netherlands Organisation for Scientific Research (NWO)
      grant 614.001.352.
      V.V.G. acknowledges support from the National Aeronautics and 
      Space Administration under grant No. 15XRP15 20140 issued 
      through the Exoplanets Research Program.
      C.W. acknowledges financial support from the University of Leeds
      \changeref{and funding from STFC (grant number ST/R000549/1).}
      This paper makes use of the following ALMA data: ADS/JAO.ALMA\#2016.1.00884.S and \#2013.1.01268.S. 
      L.I.C. acknowledges the support of NASA through Hubble Fellowship grant 
      HST-HF2-51356.001-A awarded by the Space Telescope Science Institute, 
      which is operated by the Association of Universities for Research 
      in Astronomy, Inc., for NASA, under contract NAS 5-26555.
      ALMA is a partnership of ESO (representing its member states), 
      NSF (USA) and NINS (Japan), together with NRC (Canada), NSC and 
      ASIAA (Taiwan), and KASI (Republic of Korea), in cooperation with 
      the Republic of Chile. The Joint ALMA Observatory is operated by 
      ESO, AUI/NRAO and NAOJ.
\end{acknowledgements}

%-------------------------------------------------------------------

\bibliographystyle{aa}
\bibliography{carney_phd_paper3}

% \newpage
% \vspace{10cm}

\begin{appendix}

  \section{Molecular abundances for different $T_{\rm ex}$}
  \label{app:A}
  
  \change{Here the disk-averaged column densities and abundances are calculated
  for CH$_3$OH and H$_2$CO in the disk around \changeref{HD 163296} and the disk around TW Hya
  for different excitation temperatures $T_{\rm ex}$.
  The method used is described in Section~\ref{sec:res_abun_uplims}. Equation~\ref{eq:col_dens}
  assumes optically thin emission and LTE excitation conditions. The excitation temperature
  $T_{\rm ex}$ is \changeref{set to} 25, 50 and 75 K, indicating different regions of the disk from 
  which the emission lines may originate. At most, differences of factors 2--3 are seen in 
  the disk-averaged column density and abundance for the values of $T_{\rm ex}$ explored here.}
  
  \begin{table*}[!ht]
  \caption{Disk-averaged column density and abundance of CH$_3$OH and H$_2$CO in HD 163296 and TW Hya for varying $T_{\rm ex}$.}
  \centering
  \label{tab:app_ch3oh_col_dens_abun}
  \resizebox{\textwidth}{!}{
  \begin{tabular}{llcccccccr}
  \hline \hline
  Object & Line & $\int I_{\nu} d\nu$\tablefootmark{$\dagger$} & $E_{\rm u}$ & log($A_{\rm ul}$) & $T_{\rm ex}$ & $n_{\rm crit}$\tablefootmark{a}\tablefootmark{b} & $N_{\rm avg}$ & ${\rm CH_3OH}/{\rm H_2}$ & ${\rm CH_3OH}/{\rm H_2CO}$\tablefootmark{$\dagger\dagger$}\\
  & & [mJy km s$^{-1}$] & [K] & [s$^{-1}$] & [K] & [cm$^{-3}$] & [cm$^{-2}$] &  &\\
  \hline
  \multicolumn{10}{c}{\textbf{CH$_3$OH}} \\ \\
  HD 163296 & CH$_3$OH 5$_{05}$--4$_{04}$ (E) & $<51$ & 47.9 & $-4.22$ & 25 & 1.6(06) & $<6.9 (12)$ & $<2.1 (-11)$  &\\
    & & & & & 50 & 1.9(06) & $<8.6 (12)$ & $<2.6 (-11)$  &\\
    & & & & & 75 & 2.1(06) & $<1.1 (13)$ & $<3.5 (-11)$  &\\ \\
  %  \hline
    & CH$_3$OH 5$_{05}$--4$_{04}$ (A) & $<51$ & 34.8 & $-4.22$ & 25 & 4.3(05) & $<4.1 (12)$ & $<1.3 (-11)$ & \\
    & & & & & 50 & 5.0(05) & $<6.6 (12)$ & $<2.0 (-11)$ & \\
    & & & & & 75 & 5.0(05) & $<9.6 (13)$ & $<3.0 (-11)$ & \\ \\
  %  \hline
    & CH$_3$OH 1$_{10}$--1$_{01}$ (A) & $<26$ & 16.9 & $-3.49$ & 25 & 4.3(07) & $<7.0 (11)$ & $<2.2 (-12)$ & \\
    & & & & & 50 & 5.6(07) & $<1.6 (12)$ & $<5.0 (-12)$ & \\
    & & & & & 75 & 6.4(07) & $<2.7 (12)$ & $<8.2 (-12)$ & \\ \\
  %  \hline
    & CH$_3$OH 2$_{11}$--2$_{02}$ (A) & $<26$ & 21.6 & $-3.49$ & 25 & 5.0(06) & $<5.0 (11)$ & $<1.6 (-12)$ & \\
    & & & & & 50 & 5.4(06) & $<1.1 (12)$ & $<3.3 (-12)$ & \\
    & & & & & 75 & 5.5(06) & $<1.7 (12)$ & $<5.2 (-12)$ & \\ \\
%   \hline \\
  TW Hya & CH$_3$OH stacked\tablefootmark{*} & 26.5 $\pm$ 2.7\tablefootmark{c} & 28.6 & $-3.49$ & 25 & 3.0(06) & $4.7 (12)$ & $1.1 (-12)$ & \\
    & & & & & 50 & 3.0(06) & $8.6 (12)$ & $2.1 (-12)$ & \\
    & & & & & 75 & 3.1(06) & $1.3 (13)$ & $3.2 (-12)$ & \\ \\
  \hline
  \multicolumn{10}{c}{\textbf{H$_2$CO}} \\ \\
%   Target & Line & $\int I_{\nu} d\nu$ & $E_{\rm u}$ & log($A_{\rm ul}$) & $T_{\rm ex}$ & $n_{\rm crit}$ & $N_{\rm avg}$ & ${\rm H_2CO}/{\rm H_2}$ & ${\rm CH_3OH}/{\rm H_2CO}$\tablefootmark{$\dagger$} \\
%   & & [mJy km s$^{-1}$] & [K] & [s$^{-1}$] & [K] & [cm$^{-3}$] & [cm$^{-2}$] &  & \\
%   \hline \\
  HD 163296 & H$_2$CO 3$_{12}$--2$_{11}$ & 890 $\pm$ 89\tablefootmark{d} & 33.4 & $-3.55$ & 25 & 5.7(06) & 2.1(12) & 6.3($-$12) & < 0.24 \\
  &  & & & & 50 & 6.2(06) & 3.0(12) & 9.2($-$12) & < 0.43 \\
  &  & & & & 75 & 6.4(06) & 4.2(12) & 1.3($-$11) & < 0.50 \\ \\
  %  & H$_2$CO 4$_{14}$--3$_{13}$ & 1550 $\pm$ 155 & 45.5 & $-3.23$ & 25 & 1.1(07) & 5.3(10) & 1.6($-$13) & < 0.02 \\
  %  &  & & & & 50 & 1.3(07) & 2.7(10) & 8.2($-$14) & < 0.03 \\
  %  &  & & & & 75 & 1.7(07) & 2.2(10) & 6.9($-$14) & < 0.03 \\  \\
%   \hline \\
  TW Hya & H$_2$CO 3$_{12}$--2$_{11}$ & 291 $\pm$ 29\tablefootmark{e} & 33.4 & $-3.55$ & 25 & 5.7(06) & 3.7(12) & 8.9($-$13) & 1.27 $\pm$ 0.13 \\
  & & & & & 50 & 6.2(06) & 5.3(12) & 1.6($-$12) & 1.62 $\pm$ 0.16 \\
  & & & & & 75 & 6.4(06) & 7.5(12) & 1.8($-$12) & 1.73 $\pm$ 0.17 \\  \\
  %  & H$_2$CO 4$_{14}$--3$_{13}$ & 1220 $\pm$ 122 & 45.5 & $-3.23$ & 25 & 1.1(07) & 2.3(11) & 5.5($-$14) & 0.11 \\
  %  & & & & & 50 & 1.3(07) & 1.1(11) & 2.8($-$14) & 0.15 \\
  %  & & & & & 75 & 1.7(07) & 9.5(10) & 2.3($-$14) & 0.17 \\  \\    
  \hline
  \end{tabular}
  }
  \tablefoot{The format $a$($b$) translates to $a \times 10^b$. Flux errors are dominated by systematic uncertainties, taken to be 10\%. \\
  \tablefoottext{$\dagger$}{Upper limits are derived  at the 3$\sigma$ level using the HD 163296 Keplerian mask (see Section~\ref{sec:res}).} \\
  \tablefoottext{$\dagger\dagger$}{Ratios are determined using the CH$_3$OH disk-integrated column 
  density from Table~\ref{tab:ch3oh_col_dens_abun}. HD 163296: based on 
  the strictest upper limit from the CH$_3$OH 2$_{11}$--2$_{02}$ (A) line.
  TW Hya: based on the stacked CH$_3$OH detection.}\\
  \tablefoottext{*}{The stacked detection consists of three CH$_3$OH transitions: 
  CH$_3$OH 2$_{11}$--2$_{02}$ (A) at 304.208 GHz, 
  CH$_3$OH 3$_{12}$--3$_{03}$ (A) at 305.472 GHz, 
  and CH$_3$OH 4$_{13}$--4$_{04}$ at 307.166 GHz. 
  Excitation parameters for the CH$_3$OH 3$_{12}$--3$_{03}$ (A) \changeref{line}
  are used to calculate column density.} \\
  \textbf{References:} \tablefoottext{a}{\citet{Rabli2010};} \tablefoottext{b}{\citet{Wiesenfeld2013};} \tablefoottext{c}{\citet{Walsh2016b};} \tablefoottext{d}{\citet{Qi2013};} \tablefoottext{e}{\citet{Oberg2017}.}
  }
  \end{table*}

\end{appendix}

\end{document}